\documentclass[aps,prx,showpacs,floatfix,twocolumn,superscriptaddress]{revtex4-2}

\usepackage{mathrsfs}
\usepackage[%
    colorlinks=true,
    pdfborder={0 0 0},
    linkcolor=red,
    citecolor = red
]{hyperref}
\usepackage{graphicx}
\usepackage{bm,amsmath,amssymb}
\usepackage{dcolumn}
\usepackage{xcolor}
\usepackage{bm}
\usepackage{amsmath,accents}
\newcommand{\myvect}[1]{\accentset{\rightharpoonup}{#1}}

\begin{document}
\title{Quantum coherent control in pulsed waveguide optomechanics}

\author{Junyin Zhang}
 \affiliation{Max Planck Institute for the Science of Light, Staudtstr. 2, 91058 Erlangen, Germany}
 \affiliation{Department of Physics, University of Science and Technology of China, 230026 Hefei, China}
\author{Changlong Zhu}
 \affiliation{Max Planck Institute for the Science of Light, Staudtstr. 2, 91058 Erlangen, Germany}
  \affiliation{Department of Physics, University of Erlangen-Nuremberg, Staudtstr. 7, 91058 Erlangen, Germany}
\author{Christian Wolff}
 \affiliation{Centre for Nano Optics, University of Southern Denmark, Campusvej 55, DK-5230, Odense M, Denmark}
\author{Birgit Stiller}
    \affiliation{Max Planck Institute for the Science of Light, Staudtstr. 2, 91058 Erlangen, Germany}
    \affiliation{Department of Physics, University of Erlangen-Nuremberg, Staudtstr. 7, 91058 Erlangen, Germany}

    \email[Correspondence email address: ]{birgit.stiller@mpl.mpg.de}

\date{\today} 

\begin{abstract}
Coherent control of traveling acoustic excitations in a waveguide system is an interesting way to manipulate and transduce classical and quantum information. So far, these interactions, often based on optomechanical resonators or Brillouin scattering, have been studied in the steady-state regime using continuous waves. However, waveguide experiments are often based on optical pump pulses which require treatment in a dynamic framework. In this paper, we present an effective Hamiltonian formalism in the dynamic regime using optical pulses that links waveguide optomechanics and cavity optomechanics, which can be used in the classical and quantum regime including quantum noise. Based on our formalism, a closed solution for coupled-mode equation under the undepleted assumption is provided and we found that the strong coupling regime is already accessible in current Brillouin waveguides by using pulses. We further investigate several possible experiments within waveguide optomechanics, including Brillouin-based coherent transfer, Brillouin cooling, and optoacoustic entanglement.
\end{abstract}

\maketitle

\section{introduction}
Photons are known as one of the most promising quantum information carriers in quantum communication, especially for long distances~\cite{Gisin2007} but also represent a major opportunity for quantum computation~\cite{doi:10.1126/science.abe8770}. However, enhancing photon-photon coupling is a challenge. Introducing optomechanical interaction is one of the possible ways to get photons more interactive and therefore mechanical systems have a profound impact on current quantum technologies. The combination of mature MEMS technology and the diversity of mechanical systems offer flexibility for transducing, delivering and manipulation of quantum information and open moreover new roads of exploring macroscopic quantum phenomena \cite{clerk2010introduction,poot2012mechanical,RevModPhys.86.1391}.
In addition to the considerable effort invested in optomechanical resonators~\cite{RevModPhys.86.1391,doi:10.1126/science.1244563}, some research has been conducted in waveguide optomechanics~\cite{PhysRevX.8.041034}, which may be a plausible platform for quantum networks~\cite{Habraken_2012} and quantum nonlinearity~\cite{PhysRevLett.119.123602} due to its broad bandwidth and integrability into existing circuitry \cite{li2008harnessing,kang2009tightly,kittlaus2018non}.

Waveguide optomechanics can rely on the interaction of optical waves with mechanical breathing modes of the transverse section of the optical waveguide or on traveling longitudinal acoustic waves or on hybrid versions of both of them. Brillouin scattering which describes a variety of these optoacoustic and optomechanical interactions has been experimentally investigated in detail in optical fibers and photonic integrated waveguides \cite{kobyakov2010stimulated,eggleton2019brillouin} and several classical coupled-mode equations~\cite{boyd2020nonlinear} have been analytically developed in the classical regime to understand those phenomena. Also quantum approaches have been studied, including Hamiltonian~\cite{J_E_Sipe_2016} and Lagrangian treatments~\cite{Laude_2015}, which propose a quantum field description for optomechanical interactions in waveguides. Treating the waveguide as an array of cavities is another way towards optomechanical waveguide theories~\cite{rakich2018quantum}. Current research on quantum regimes in an analytical way however focuses on steady-state behaviors and treat continuous-wave (CW) interactions. Studying dynamic processes including optical pump pulses involve challenges such as the contradiction between finite control length and the infinite expanded nature of phonons. This makes it challenging to analyze the time-dependent quantum evolution in optoacoustic processes stimulated by a pulsed pump such as Brillouin-based memory \cite{zhu2007stored,merklein2017chip,stiller2020coherently}.
    \begin{figure}
    \centering
    \includegraphics[width=0.45\textwidth]{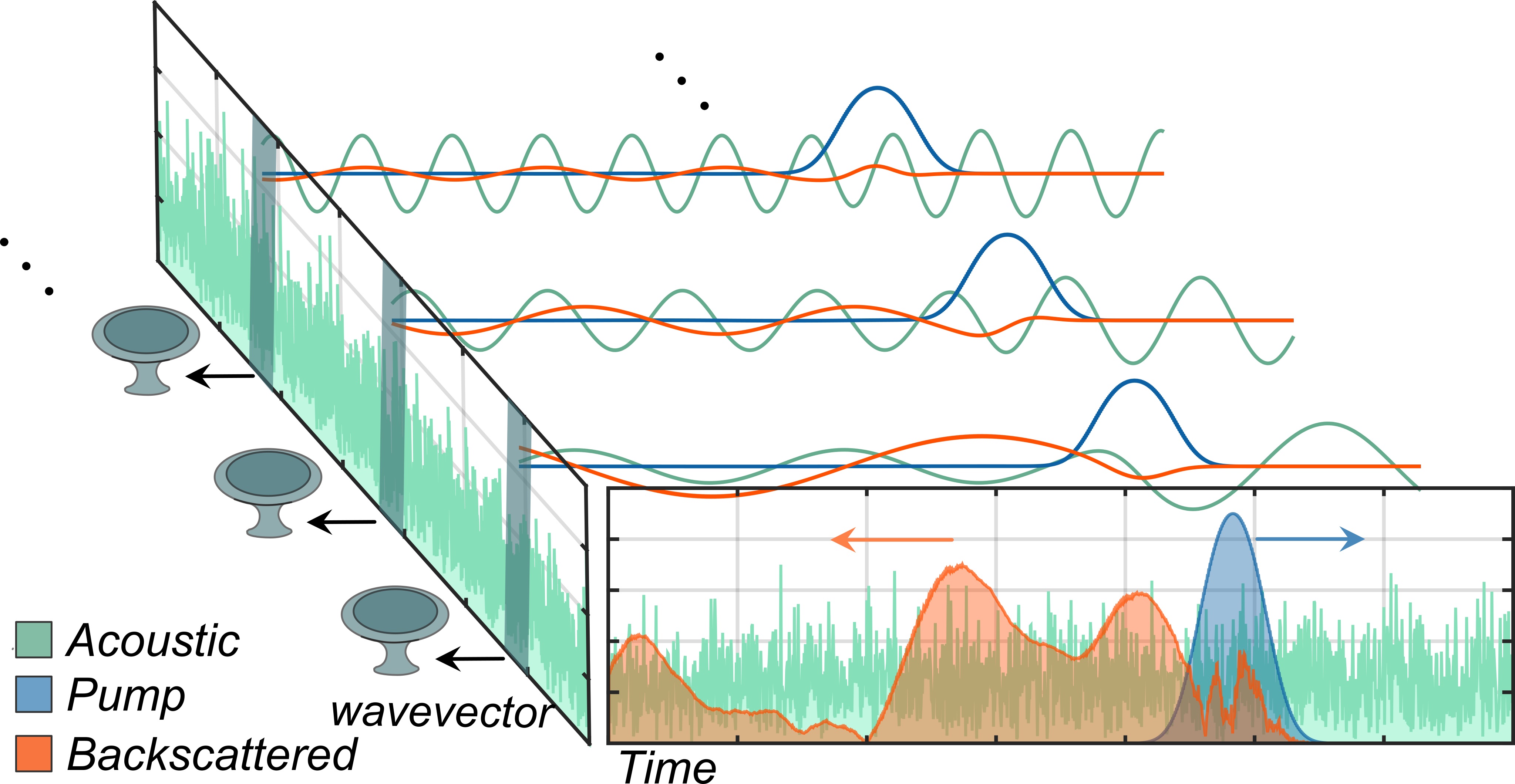}
    \caption{Principle: the backward Brillouin interaction can be separted in different wavevector channels. In each channel, the backscattered nature contributes to a cavity-like localized interaction.
    }
    \label{fig:optomechanics_general}
    \end{figure}

In this work, we formulate a framework in which backscattered Brillouin scattering in a waveguide is treated as a continous cavity array in the momentum space (FIG.~\ref{fig:optomechanics_general}).

We introduce co-moving coordinates that move together with the optical pump pulse which enables a simple formulation of the pulsed dynamical regime, in which propagating optical pulses interact with traveling acoustic waves, respectively. Under the assumption of sufficiently short undepleted pump pulses, the framework maps the dynamic pulsed case in a waveguide into a cavity-like system which greatly simplifies the treatment of different scenarios of coherent control. We find that due to the larger depletion threshold for shorter pulses, the strong coupling regime can be achieved using current platforms in the pulsed regime. With this framework, we explore several challenging problems in backward Brillouin scattering in waveguide systems, such as coherent transfer between photons and phonons, Brillouin cooling, and entanglement in Brillouin waveguide systems.

First, we analytically demonstrate that by delicately controlling the pump pulse length, coherent transfer and Brillouin cooling with high efficienty using the backward Brillouin process is possible. More specifically, we use an anti-Stokes Brillouin process in the backward regime to show that the quantum state can be transferred between photons and acoustic phonons. {This transfer can be used for cooling the acoustic phonons in longer fibers with higher efficiency than those based on continuous waves \cite{ PhysRevX.8.041034,chen2016brillouin}.} Note that previous works were mostly based on forwarding Brillouin scattering, which is related to transverse mechanical vibrations and presents a different operation regime \cite{ PhysRevX.8.041034,chen2016brillouin}.
Secondly, we analytically show that entangled pair generation is possible by using the backward Brillouin scattering Stokes process. Our evaluation suggests that these regimes can be attained by existing waveguide systems such as chalcogenide fibers and nano-scale waveguides.

   The paper is organized as follows: In Section.\ref{section:Effective Hamiltonian Formulation} we briefly summarize the conventional backward Brillouin interaction and then present the effective Hamiltonian formalism, which is the main result of this paper. In Section.\ref{Section:Application}, we investigate three challenging problems in waveguide optomechanics using our formalism: coherent transfer, Brillouin cooling, and entangled pair generation. In the last Section.\ref{Conclusion}, we summarize our result and discuss several open questions in waveguide optomechanics.

\section{Effective Hamiltonian Formulation}
\label{section:Effective Hamiltonian Formulation}
\subsection{Waveguide Optomechanical System}

We consider an optomechanical waveguide system,
which allows the guidance of both electromagnetic and acoustic waves with different wave vectors and in different spatial modes. A typical optomechanical interaction in such a waveguide system treats a mechanical oscillation with frequency $\Omega({k})$, a light field with optical frequency $\omega({k})$ and the optomechanical coupling $g_{0}$ which refers to the coupling between two photons with wave vectors $k_S, k_p$ ($p,q$) and one acoustic phonon with wave vector $q$($k$). The optomechanical coupling can originate from different physical processes such as electrostriction\cite{Wolff:21} and radiation pressure\cite{PhysRevLett.107.063602}. Considering the three-wave-mixing optomechanical coupling (usually the dominants one),
the system can be described by the Hamiltonian~\cite{J_E_Sipe_2016,rakich2018quantum}:
\begin{equation}
    \begin{aligned}
        H &= \int_{-\infty}^{+\infty}\text{d}q\ \hbar \omega\left(q\right)a_q^\dagger a_q
    + \int_{-\infty}^{+\infty}\text{d}q\ \hbar \Omega\left(q\right)b_q^\dagger b_q \\
    &\quad\ + \hbar \iint_{-\infty}^{+\infty}\text{d}q\text{d}p\ \left(g_{p,q} a_{p+q}^\dagger a_p b_q + h.c.\right)\ ,
    \end{aligned}
        \label{eq:general_hamiltonian}
\end{equation}
where $a_{k}$ and $b_{k}$ are annihilation operators of the electromagnetic and mechanical modes. {For the interactions within the narrow frequency band of interest, the coupling factor $g_{p,q}$ can be approximated by a coupling constant $g_0$.}

This Hamiltonian can be derived from combining elastic theory and Maxwell's equations by introducing an optomechanical coupling as the interaction part ~\cite{J_E_Sipe_2016}. To treat this quantum mechanically, it is then quantized on the normal modes. The first two/three terms are the energy for photons and acoustic phonons (the free Hamiltonian part $H_0$) and the last two terms are the interaction Hamiltonian $H_I$.

\begin{figure}
\centering
\includegraphics[width=0.45\textwidth]{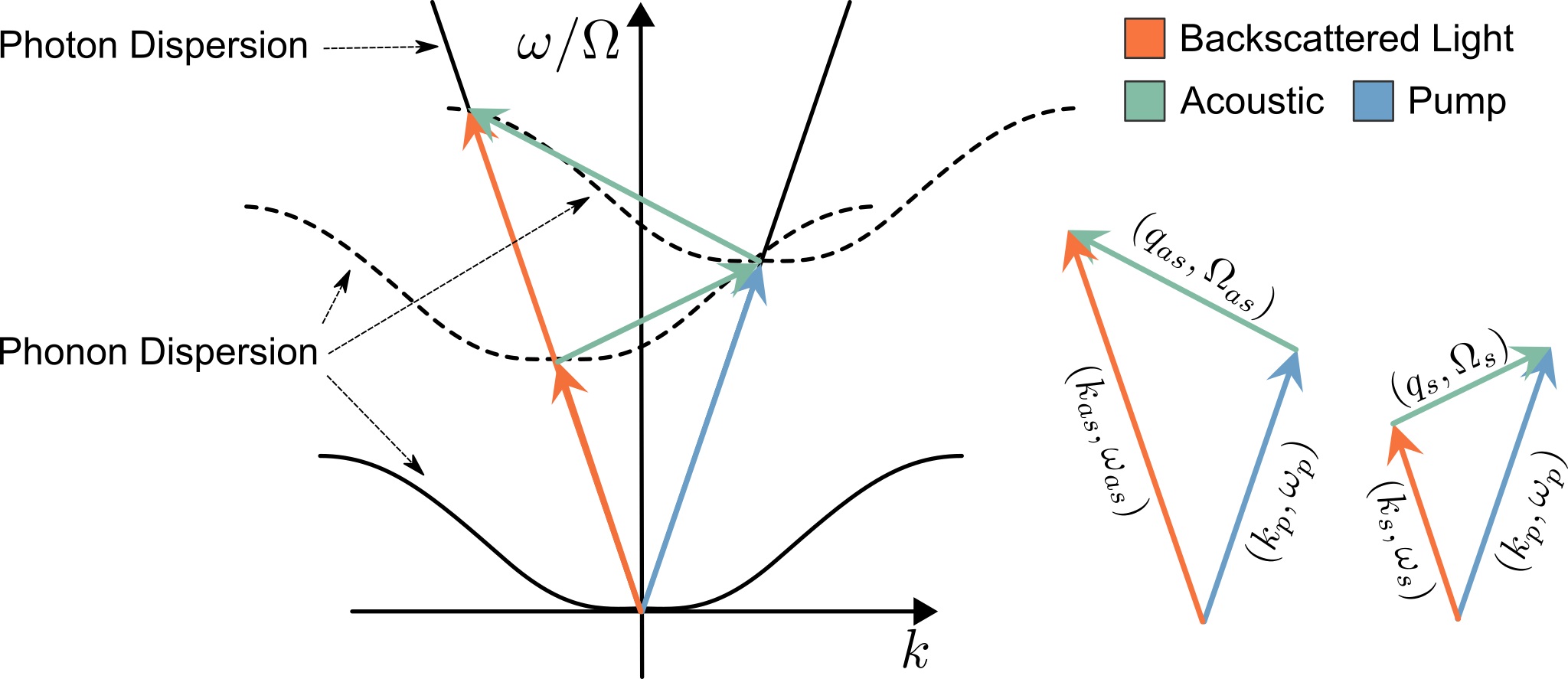}
\caption{The phase-matching diagram. The black corn and black curve refer to the dispersion relation of photons and acoustic phonons. The black dotted line is the copy of the phonon dispersion relation. For the Stokes process, the phase-matching condition can only be satisfied when the pump photon emits a forward traveling phonon (the green arrow pointing to the top right corner). In contrast, the anti-Stokes process can only be stimulated when absorbing a backward traveling phonon(the green arrow pointing to the top left corner). }
\label{fig:phase_matching}
\end{figure}

The optomechanical interactions in the waveguide are constrained by the phase-matching condition. As shown in Fig.\ref{fig:phase_matching}, there are two points~\cite{Wolff:21} where the phase-matching condition is satisfied when the waveguide is pumped by an optical field with wavevector $k_p$ and frequency $\omega(k_p)$. These two points refer to the Stokes process/anti-Stokes process with photon wavevector $k_{s/as}$, photon frequency $\omega(k_{s/as})$, phonon wavevector $q_{s/as}$ and phonon frequency $\Omega(q_{s/as})$. Since the two kinds of phonons corresponding to the Stokes and anti-Stokes process are separated in directions, we can selectively apply the Stokes process or Anti-Stokes process on the directional traveling phonons by choosing the pump pulse direction.

The wave package operators are defined by integrating the wavevectors around the phase-matching points~\cite{J_E_Sipe_2016}:
\begin{equation}
\begin{aligned}
    a_{p/as/s}(z,t) &=\frac{1}{\sqrt{2\pi}} \int_{-\infty}^{+\infty}\text{d}k\\
    &\quad\  a_k \mathrm{e}^{-\mathrm{i}(k-k_{p/as/s})z}\mathrm{e}^{\mathrm{i}\omega(k_{p/as/s})t}\ , \\
    b_{as/s}(z,t) &= \frac{1}{\sqrt{2\pi}}\int_{-\infty}^{+\infty}\text{d}q\\
    &\quad\  b_q \mathrm{e}^{-\mathrm{i}(q-q_{s/as})z}\mathrm{e}^{\mathrm{i}\omega(q_{s/as})t}\ , \\
\end{aligned}
\end{equation}
where the first equation corresponds to the optical waves and the second equation to the acoustic wave. The fiber is placed on the $z$ axis in the lab frame and the pump wave is assumed to be propagating along the positive $z$ axis. It can be verified that the wave package operators and the corresponding Hermite conjugates are well-defined quantum operators that preserve the commutation relations. Utilizing the wave package operators, the interaction part $H_I$ can be written as the products in both Stokes process ($H_{I, S}$) and the anti-Stokes ($H_{I, AS}$) process:
\begin{equation}
    \begin{aligned}
    &H_{I,s} = g_0\hbar \int_{-\infty}^{+\infty}\text{d} z\ a_p^\dagger(z)a_{s}(z)b_s(z) + \text{h.c.}\ , \\
    &H_{I,as} = g_0\hbar \int_{-\infty}^{+\infty}\text{d} z\ a_p(z)a_{as}^\dagger(z)b_{as}(z) + \text{h.c.}\ .
    \end{aligned}
    \label{eq:the_Hamiltonian}
\end{equation}

From the Heisenberg equation of the optomechanical Hamiltonian, we can derive the approximated motion equation for $a_{p/s/as}(z,t)$ and $b_{as/s}(z,t)$, which are the coupled-mode equations for the stimulated Brillouin scattering process. Here, we write the coupled-mode equations for the backward Brillouin process which will be the important process throughout this paper.

For the Stokes process, the equations are:
\begin{equation}
\begin{aligned}
\partial_t a_p + c_g\partial_z a_p &= -\mathrm{i}g_0 a_s b_s - \gamma/2\  a_p \ ,\\
\partial_t a_s - c_g\partial_z a_s &= -\mathrm{i} g_0 a_p b_s^\dagger - \gamma/2\  a_s\ ,\\
\partial_t b_s + u_g\partial_z b_s &= -\mathrm{i} g_0 a_p a_s^\dagger - \Gamma/2\  b_s + \sqrt{\Gamma} {\xi}\ .
\end{aligned}
\label{eq:Stokes_original}
\end{equation}

For the anti-Stokes process, the equations are:
\begin{equation}
\begin{aligned}
&\partial_t a_p + c_g\partial_z a_p = -\mathrm{i} g_0 a_{as} b_{as}^\dagger - \gamma/2\  a_p\ ,\\
&\partial_t a_{as} - c_g\partial_z a_{as} = -\mathrm{i} g_0 a_p b_{as} - \gamma/2\  a_{as}\ ,\\
&\partial_t b_{as} - u_g \partial_z b_{as} = -\mathrm{i} g_0 a_p^\dagger b_{as} -\Gamma/2\  b_{as} + \sqrt{\Gamma} {\xi}\ .
\end{aligned}
\label{eq:Anti-Stokes_original}
\end{equation}

The $a_p, a_s, a_{as}$ refer to the optical wave packets for the pump wave, Stokes wave, and the anti-Stokes wave respectively. The $b_s$ and $b_{as}$ refer to the acoustic wave packets related to the two processes. The optical excitations and the acoustic excitations travel in the fiber in different group velocities described by $c_g, u_g$ and suffer a dissipation rate with $\gamma, \Gamma$. Since the acoustic excitations travel much slower than the optical ones, therefore we can omit this effect by setting $u_g=0$ in the following discussion. The thermal noise in the acoustic fields is taken into consideration with the Langevin term $\sqrt{\Gamma}\tilde{\xi}$, where $n_{th}$ is the averaged thermal phonon number at the given temperature as:
\begin{equation}
    n_{th} = \frac{1}{\mathrm{e}^{\epsilon_p/k_B T} - 1}\ ,
\end{equation}
where $\epsilon_p$ is the energy of a single phonon. $\tilde{\xi}=\tilde{\xi}(z,t)$ is a random function:
\begin{equation}
\begin{aligned}
    \langle {\xi}(z_1,t_1){\xi}^\dagger(z_2,t_2)\rangle &=n_{th} \delta(z_1-z_2,t_1-t_2)\ ,\\
    [{\xi}(z_1,t_1),{\xi}^\dagger(z_2,t_2)] &= \delta(z_1-z_2,t_1-t_2)\ .
\end{aligned}
\end{equation}

The coupled-mode equation above can be used for numerical simulations. However, finding an analytical solution for the dynamical cases of optical and acoustic pulses and investigating analytically quantum phenomena are both not straight forwards with this set of equations. In the following sections, we will show that the coupled mode equations can be exactly solved under the undepleted assumption.

\subsection{The undepleted assumption}
In the following discussions, we consider the undepleted case. The undepleted assumption refers to the condition where the waveform function of the pump light $a_p(z,t)$ remains unchanged during the scattering process. This assumption is valid for quantum Brillouin experiments, where the amplitude of the quantum-level acoustic field and backscattered field like Stokes and anti-Stokes wave are too small to deplete the pump significantly.

Defining:
\begin{equation}
    g(z,t) = g_0 \langle a_p(z,t)\rangle
    \label{Eq:Effective Coupling}\ .
\end{equation}
Since the pump waveform $a(z,t)$ remained unchanged during the propagation, the $g(z,t)$ would remained unchanged too:
\begin{equation}
    a(z,t) = a(0,t-z/c_g) \Longleftrightarrow g(z,t) = g(0,t-z/c_g)\ .
\end{equation}
Under the undepleted assumption, the first equation in the coupled mode equations, which refers to the pump dynamics, can therefore be omitted. For example, the coupled mode equation for the Stokes process can be linearized as:
\begin{equation}
    \begin{aligned}
    &\partial_t a_s - c\partial_z a_s = -\mathrm{i} g b_s^\dagger - \gamma/2\  a_s\ ,\\
    &\partial_t b_s= -\mathrm{i} g a_s^\dagger - \Gamma/2\  b_s + \sqrt{\Gamma} {\xi}\ .
    \end{aligned}
\end{equation}

Under the undepleted assumption, the equations are now linearized. The effective coupling strength $g(z,t)$ describes the coupling between the acoustic and optical fields, which is tunable by changing the pump power. This tunable coupling strength enables us to control the acoustic phonons and photons traveling in the waveguide system coherently. Furthermore, as discussed later, we will show that the strong coupling conditions $g>\Gamma$ are possible by applying strong short pump pulses under the pulse Brillouin threshold. The detailed strong coupling condition and the pulsed threshold are discussed later in Sec.~\ref{sect:undepleted_conditions}.

\subsection{Boundary Value Problem}
Considering a fiber of length $L$, we assume that there is no light in the fiber at $t=0$ and all the light waves are input at either the end at $z=0$ or the end at $z=L$ of the fiber at $t>0$. To determine to whole evolution of all wave packets functions in the fiber at $t>0, 0<z<L$, we need to know: 1. the initial acoustic states $b_{s/as}(z,t=0)$; 2. the input pump waveform $a_p(z=0,t)$, when linearized, this terms can be fully described by $g(z=0,t)$; 3. the input backscattered waveform $a_{s/as}(z=L,t)$; 4. the detailed form of the noise function $\tilde{\xi}(z,t)$. If we only care about the statistical result, the detailed form of the noise function is not needed.

\begin{figure}
\centering
\includegraphics[width=0.35\textwidth]{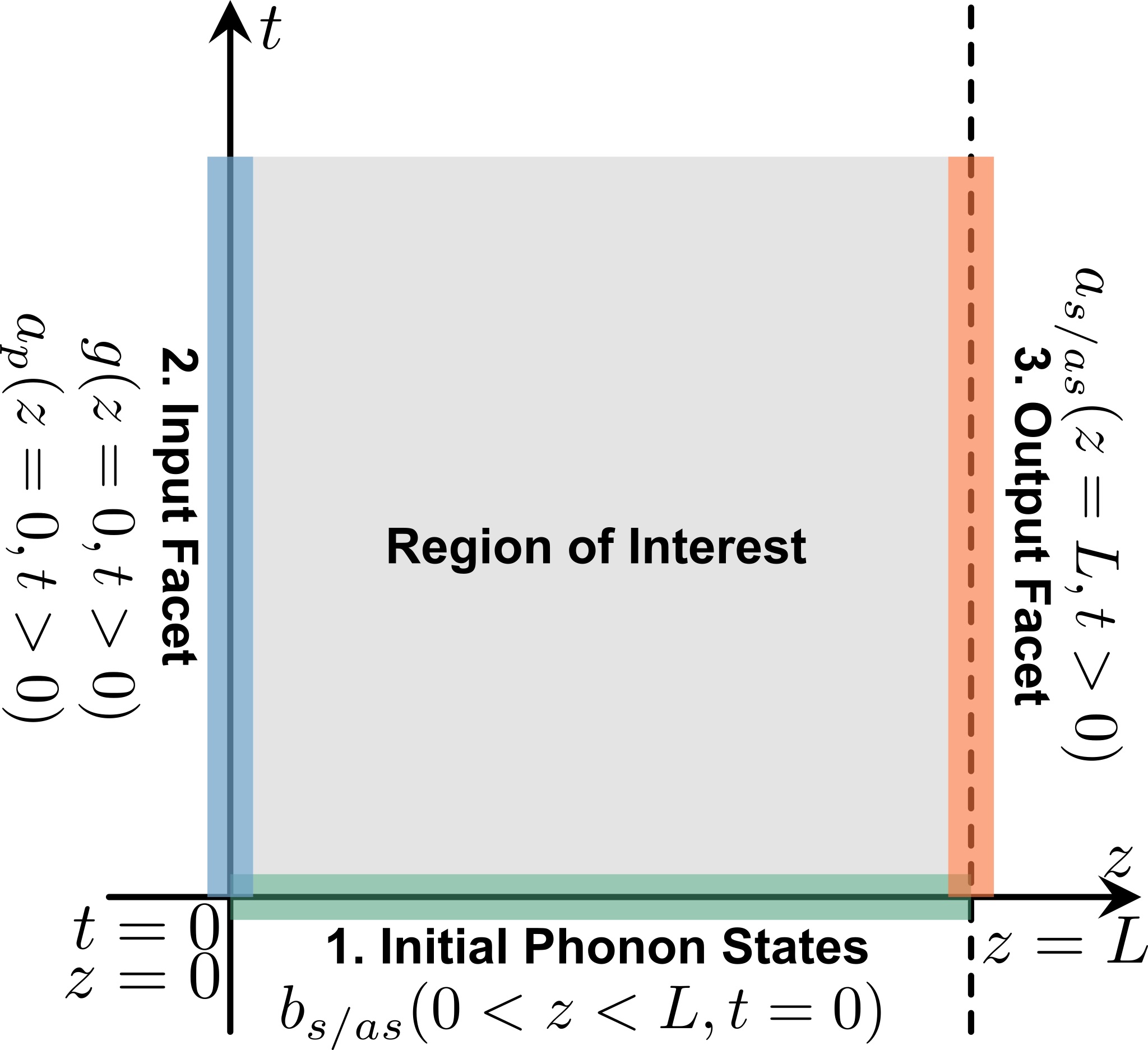}
\caption{The region of interest in backscattered Brillouin scattering problems. Besides noise, three boundary conditions need to be specified for both Stokes process and the anti-Stokes process so that the coupled mode equation can be mathematically complete: 1. the initial phonon states at $t=0$. 2. the input pump waveform at $z=0$. 3. the input backscattered waveform at $z=L$.}
\label{fig:illa_1}
\end{figure}

The scattering process can be described in a space-time diagram like the following, in this space-time diagram, the above requirements all appear at the boundaries, as shown in FIG.~\ref{fig:illa_1}.

\subsection{Exact Solution under undepleted assumption}

By defining the coordinates transformation:
\begin{equation}
    \begin{aligned}
    \eta &= t-z/c\ ,
    & \quad
    \tau &= t\ ,
    \end{aligned}
\end{equation}
the distribution coupling strength $g(z,t)$ can be directly related to the boundary conditions $a_p(z=0,t)$):
\begin{equation}
    g(z,t) = g(0,t-z/c_g) = g(\eta)\ .
\end{equation}
Take the Stokes process as an example, the linearized coupled mode equations for the Stokes process can be further written as:
\begin{equation}
\begin{aligned}
    (\partial_\tau + 2\partial_\eta) a_s &= -\mathrm{i} g b_s^\dagger - \gamma/2\  a_s\ ,\\
    (\partial_\tau + \partial_\eta) b_s^\dagger &=  \mathrm{i} g a_s - \Gamma/2\  b_s^\dagger + \sqrt{n_{th}\Gamma}{\xi}^\dagger\ .
\end{aligned}
\end{equation}
Since $g$ only depends on $\eta$, the variables in the above equation can be seperated by performing a Fourier transformation on variable $\tau$:
\begin{equation}
    \begin{aligned}
        \tilde{a}_s(\Delta,\eta) = \frac{c_g}{L}\int_0^{L/c_g}\text{d}\tau\ a_s(\tau,\eta) \mathrm{e}^{-\mathrm{i}c_g\Delta\tau}\ ,\\
        \tilde{b}_s^\dagger(\Delta,\eta) = \frac{c_g}{L}\int_0^{L/c_g}\text{d}\tau\ b_s^\dagger(\tau,\eta) \mathrm{e}^{-\mathrm{i}c_g\Delta\tau}\ ,\\
        \tilde{{\xi}}^\dagger(\Delta,\eta) = \frac{c_g}{L}\int_0^{L/c_g}\text{d}\tau\ \tilde{\xi}^\dagger(\tau,\eta) \mathrm{e}^{-\mathrm{i}c_g\Delta\tau}\ .\\
    \end{aligned}
\end{equation}
Then the equation can be written as the following Langevin form:
\begin{equation}
\begin{aligned}
    \partial_{\eta}
    \begin{pmatrix}
    \tilde{a}_s\\
    \tilde{b}_s^\dagger
    \end{pmatrix}
    &=
    \begin{pmatrix}
    -\tfrac{\gamma + 2 \mathrm{i}c_g\Delta}{4} & -\tfrac{\mathrm{i}g}{2}\\
    \mathrm{i}g & -\tfrac{2 \mathrm{i}c_g\Delta + \Gamma}{2}
    \end{pmatrix}
    \begin{pmatrix}
    \tilde{a}_s\\
    \tilde{b}_s^\dagger
    \end{pmatrix}
    +
    \begin{pmatrix}
    0\\
    \sqrt{\Gamma }\tilde{{\xi}}^\dagger
    \end{pmatrix}\ .
\end{aligned}
\label{Eq:Stokes_matrix}
\end{equation}

 It has to be noted that the closed solution for anti-Stokes process can be obtained in the same way. For the anti-Stokes process, a similar Langevin form can be obtained:
\begin{equation}
\begin{aligned}
    \partial_{\eta}
    \begin{pmatrix}
    \tilde{a}_{as}\\
    \tilde{b}_{as}
    \end{pmatrix}
    &=
    \begin{pmatrix}
    -\tfrac{\gamma + 2 \mathrm{i}c_g\Delta}{4} & -\tfrac{\mathrm{i}g}{2}\\
    -\mathrm{i}g & -\tfrac{2\mathrm{i}c_g\Delta + \Gamma}{2}
    \end{pmatrix}
    \begin{pmatrix}
    \tilde{a}_{as}\\
    \tilde{b}_{as}
    \end{pmatrix}
    +
    \begin{pmatrix}
    0\\
    \sqrt{\Gamma}\tilde{{\xi}}
    \end{pmatrix}\ .
\end{aligned}
\label{eq:anti-Stokes matrix}
\end{equation}

{The matrices in the above equations can be  made Hermitian by variable substitution: $\tilde{A} = \frac{\sqrt{2}}{2}\tilde{a},\ \tilde{B} = \tilde{b}$.} {The initial phonon states and backward laser injections (Stokes and anti-Stokes part) are included in the initial conditions at $\eta=0$. The above equations, describing the evolution of each Fourier component $\Delta$, are similar to the Langevin equation in optomechanics cavities. The only difference is the time-evolution in optomechanics cavities is now replaced by the $\eta$-evolution on a co-propagating framework. Each Fourier component with off-resonance variable $\Delta$ can be written with the corresponding equation of $\eta$ evolution, and the equations are independent of each other. This separability indicates a new viewpoint of waveguide optomechanics that by separating the interactions to different frequency/wavevector channels, cavity-like behavior can be recovered. The detail of this similarity is depicted in FIG.~\ref{fig:optomechanics_general} and is discussed in detail in Sec.\ref{physics_understanding}.}\

\begin{figure}
\centering
\includegraphics[width=0.40\textwidth]{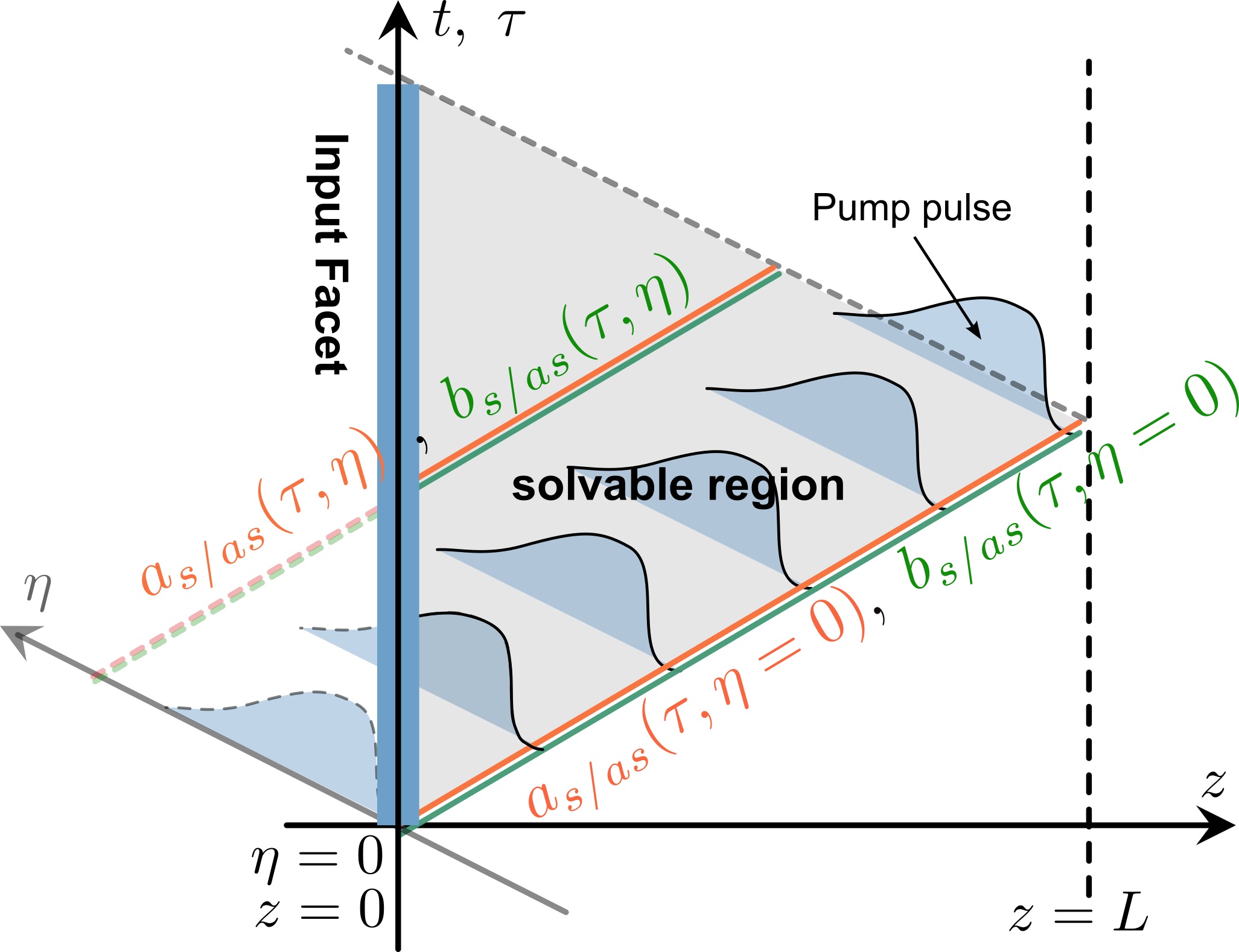}
\caption{The exact solvable region. Under the undepleted assumptions, seperation of variables could be applied to the coupled mode equation after variable transformation $(z,t)\to (\tau,\eta)$. In this case, the exact solution of the scattering process can be obtained in the gray triangle region shown in this figure. To obtain the exact solution in this region, two boundary conditions are needed: 1. the phonon and backscattered waveform at $\eta=0$: $a_{s/as}(\tau,\eta=0),\ b_{s/as}(\tau,\eta=0)$. 2. the coupling strength at every $\eta$: $g(\eta)=g_0 a_p(0,t=\eta)$.}
\label{fig:illa_2}
\end{figure}

The Langevin equations above are exactly solvable, which means that once we obtained the boundary conditions at $\eta = 0$, then the exact solutions can be obtained in the solvable region as shown in FIG.~\ref{fig:illa_2}.
As with the conventional Langevin equations, a Green function, also called the time evolution operator, can be introduced to obtained the exact solution. The Langevin equation in a general matrix form reads:
\begin{equation}
    \frac{d\myvect{M}(\eta)}{d\eta} = \mathbf{P}(\eta) \myvect{M}(\eta) + \myvect{R}(\eta)\ .
\end{equation}
By introducing the Green function:
\begin{equation}
    \mathbf{G}(\eta_2,\eta_1) = \mathcal{T}\left\{\exp{\int_{\eta_1}^{\eta_2}\text{d}\upsilon\ \mathbf{P}(\upsilon)}\right\}\ .
\end{equation}
The $\mathcal{T}\{\dots\}$ is the time-ordering operator to ensures the exponential is time-ordered, which in context means $\eta$-ordered: any product of $\mathbf{P}(\upsilon)$ that occurs in the expansion of the exponential must be ordered such that the value of $\upsilon$ is increasing from right to left of the product. The solution can be written as:
\begin{equation}
    \myvect{M}(\eta) = \mathbf{G}(\eta,0)\myvect{M}(0) + \int_0^\eta \text{d}\upsilon\  \mathbf{G}(\eta,\upsilon)\myvect{R}(\upsilon)\ .
\end{equation}

As shown in Eq.~(\ref{Eq:Effective Coupling}), the pump laser can be fully described as a modification of the effective coupling. Under the undepleted assumption, the pump pulse travels along the fiber without changing its waveform. Therefore all points on the fiber experience the same effective coupling waveform. In our formalism (Eq.~(\ref{Eq:Stokes_matrix}) and Eq.~(\ref{eq:anti-Stokes matrix})), this can be fully described by $g(z,t) = g(0,t-z/c) = g_0 \langle a_p(0,t-z/c)\rangle$ using the pulse waveform at the input port. Therefore we can claim that all undepleted situation has been solved.

To demonstrate the formalism developed here, we consider a rectangular pump shape, where:
\begin{equation}
    g(\eta) = g\Theta(\eta)\Theta(T-\eta)\ .
    \label{eq:RectanglePulse}
\end{equation}
In this case, the solution can be written as (when $t\leq T$):
\begin{equation}
    \myvect{M}(\eta) = \mathbf{G}(\eta)\myvect{M}(0) + \int_0^\eta \text{d}\upsilon\  \mathbf{G}(\eta-\upsilon)\myvect{R}(\upsilon)\ .
    \label{Eq:solution_of_langevin}
\end{equation}
with
\begin{equation}
    \mathbf{G}(\eta) = \exp(\mathbf{P}\eta)\ .
    \label{eq:propagator_G}
\end{equation}

\subsection{Physics Meaning: Co-moving Conditions}
\label{physics_understanding}
In this section, we show that the physics behind the variable-separated equations reveals a similarity between waveguide systems and cavity systems. Using Fourier transformation, the solving process can be separated into different $\Delta$ Fourier basis functions, {as shown in Eq.~(\ref{Eq:Stokes_matrix}) and Eq.~(\ref{eq:anti-Stokes matrix})}. Consider the $\Delta_0$ channel after Fourier transformation at the $\eta-\tau$ boundary $\eta=0$: $\tilde{a}(\tilde{b})_s(\Delta_0,\eta=0)=a_0(b_0)$. When the noise and dissipation is omitted, this boundary condition in the $\eta-\tau$ coordinates can be related to that in the $z-t$ coordinates, which is $a(b)_s(z,t=0)$. Since the pump $g(\eta)=0$ when $\eta<0$ here, we have:

\begin{equation}
    \begin{aligned}
        &\tilde{a}_s(\Delta,\eta=0) = a_0\delta(\Delta-\Delta_0)\\\ &\quad\quad\Longleftrightarrow\   a_s(z,t=0) = a_0 \mathrm{e}^{\mathrm{i}\Delta z/2}\ ,\\
        &\tilde{b}_s^\dagger(\Delta,\eta=0) = b_0^\dagger\delta(\Delta-\Delta_0)\\ &\quad\quad\Longleftrightarrow\   b_s^\dagger(z,t=0) = b_0^\dagger \mathrm{e}^{\mathrm{i}\Delta z}\ ,\\
    \end{aligned}
\label{eq:special_initial}
\end{equation}
which means that, under the undepleted assumptions, the interaction between different wavevector pairs can be seperated, as shown in FIG.~\ref{fig:optomechanics_general}.
The phase factor on $a_s$ is different by a factor of $2$ with $b_s$, which is a result of its opposite traveling direction to the pump pulse.
 One of the main differences between optomechanical waveguides and optomechanical cavities is the Hilbert space. In optomechanical resonators, the phonon states and the photons states are discrete. In waveguides, the phonon and photon spectrum is continuous. Therefore, we have to consider a spectrum-dependent interaction in the optomechanical waveguide. In our case, the interaction is only significant near the phase matching point, and $\Delta$ here is the wavevector deviation from the phase matching point.

\begin{figure}
\centering
\includegraphics[width=0.40\textwidth]{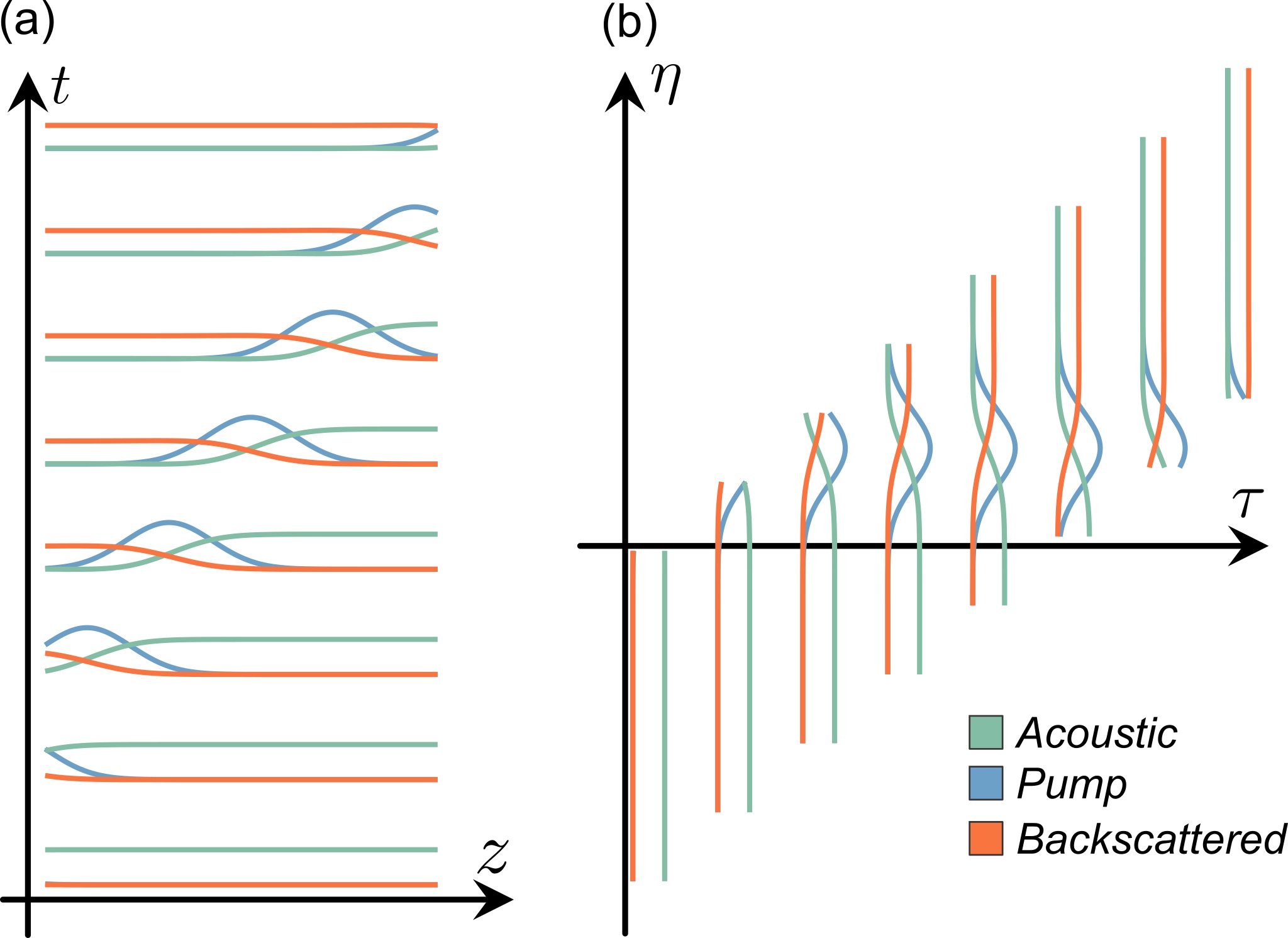}
\label{fig:co_moving}
\caption{The physics behind the solutions: co-moving conditions. If we omits the noise and depletion, a co-moving condition Eq.~(\ref{eq:co_moving_conditions}) can be derived once the plane-wave like initial conditions in Eq.~(\ref{eq:special_initial}) is satisfied: (a) In the $z-t$ coordinates, or the lab frame,  both the backscattered light and the acoustic waveform are co-moving with the pump (b) In the $\eta-\tau$ coordinates, all the waveforms are constant along the $\tau$ axis. The phase factor is omitted in both (a) and (b). This co-moving condition means that in the co-moving frame of the pump pulse, the waveform of both acoustic and backscattered waves will remain unchanged except for a phase factor $\mathrm{e}^{\mathrm{i}\Delta z}$. This makes the non-localized interaction process in the waveguides similar to the localized interaction in the optical cavities, and this is the physics reason why we can treat waveguides using similar methods in the cavities.}
\label{fig:co_moving_ridge}
\end{figure}

Furthurmore, in this case, from the Langevin equations we have:
\begin{equation}
    \begin{aligned}
    a_s(\tau,\eta) &= a_s(0,\eta)\mathrm{e}^{\mathrm{i}c_g\Delta\tau}\ ,\\
    b_s^\dagger(\tau,\eta) &= b_s^\dagger(0,\eta)\mathrm{e}^{\mathrm{i}c_g\Delta\tau}\ .\\
    \end{aligned}
\end{equation}
Using the space and time variables in the lab frame $(z,t)$, the above equations result in:
\begin{equation}
    \begin{aligned}
    \forall x>0&:\\
    &\quad a_s(z+c_gx,t+x) = a_s(z,t)\mathrm{e}^{\mathrm{i}c_g\Delta x}\ ,\\
    &\quad b_s^\dagger(z+c_gx,t+x) = b_s^\dagger(z,t)\mathrm{e}^{\mathrm{i}c_g\Delta x}\ .
    \end{aligned}
    \label{eq:co_moving_conditions}
\end{equation}
This condition means that the waveform of both the acoustic field and the backscattered optical field remain unchanged in the co-moving frame along with the propagating pump except for an additional phase factor $\mathrm{e}^{\mathrm{i}cq\tau}$, as shown in FIG.~\ref{fig:co_moving_ridge}. This co-moving frame in a special case $\Delta=0$ is also discussed in Ref.~\cite{Keaton:14}. This turns the non-localized interaction in waveguides into a cavity-like localized interaction in the co-moving frame.  Figure.\ref{fig:optomechanics_general} depicts the consequence of this assumption: The whole scattering process in the waveguide can be separated into different frequency channels. In each of the channels, translational invariances hold, and that enables us to build a mathematical framework to connect the waveguide optomechanics and the cavity optomechanics.

This result indicates that the pump pulse length can be used to control the evolution time for phonons and photons in waveguides. Exactly controlling the interaction time enables us to control the optoacoustic interaction coherently which makes the dynamic regime more interesting for coherent control applications, in contrast to the steady-state regime. In the following sections, we will discuss some predictions derived from this formalism.

\subsection{Waveguide Optomechanics in the Perspective of Cavity Optomechanics: with Backward Brillouin Scattering as an Example }

\subsubsection{The beam-splitter-like and down-conversion-like interaction}
{The Brillouin interaction Hamiltonian is described in Eq.~(\ref{eq:the_Hamiltonian}). In the case of linearizing the interaction using undepleted pump assumption, the interaction Stokes Hamiltonian $H_{I,S}$ describes a down-conversion-like process between the Stokes photons and the phonons, while the anti-Stokes part $H_{I,AS}$ describes a beam-splitter-like  process, correspondingly \cite{RevModPhys.86.1391}.}

{In terms of cavity optomechanics, the beam-splitter-like interaction (the anti-Stokes process here) describes a state transfer between the anti-Stokes photons and phonons. Such converting process is the Rabi-oscillation and can be used to achieve coherent transfer. The area dependency rule in Brillouin memory is exactly the result of area dependency in Rabi-oscillation \cite{Dong:15}, as explained later.}

{As for the down-conversion-like interaction (the Stokes process here), which corresponds to an amplification process in the Stokes photons and phonons. This parametric amplification is the reason why the Stokes light intensity is usually much larger than the anti-Stokes light intensity in conventional Brillouin experiments. This parametric amplification process in Stokes interaction can also be used to generate entangled pairs between photons and phonons. }

\subsubsection{The Brillouin gain and the strong coupling  regime}

In this section, we relate our approach to commonly used experimental parameters in backward stimulated Brillouin scattering (SBS) experiments.

In SBS, the Stokes process is dominant. We consider the steady state in which both $\frac{\partial a_{s}}{\partial t}$ and $\frac{\partial b}{\partial t}$ equal to zero in Eq.(\ref{eq:Stokes_original}) and introduce the acoustic dissipation rate $\Gamma$. Then we get:
\begin{equation}
    \frac{\partial}{\partial z} (a_S^\dagger a_s) = -\frac{4g_0^2 a_p^\dagger a_p}{\Gamma c_g} a_S^\dagger a_S\ .
\end{equation}
In SBS generated by a continuous pump, it holds: $\frac{\partial}{\partial z} (a_S^\dagger a_s) = -G P a_S^\dagger a_S$. The $G$ refers to the Brillouin gain whose unit is $[m^{-1}W^{-1}]$ and the $P$ refers to the pump power, whose unit is $[W]$. Therefore the effective gain we introduce can be obtained from the pump power directly~\cite{boyd2020nonlinear}:
\begin{equation}
    |g| = \sqrt{\frac{GP\Gamma c_g}{4}}\ .
    \label{eq:coupling}
\end{equation}
A dimensionless effective coupling ratio $\frac{g}{\Gamma}$ can be introduced by utilizing Eq.(\ref{eq:coupling}):
\begin{equation}
    \frac{|g|}{\Gamma}= \sqrt{\frac{GPc_g}{4\Gamma}}\ .
    \label{eq:Coupling ratio}
\end{equation}
Coherent control is only possible when the effective coupling ratio is larger than one: ${g}/{\Gamma}>1$. As shown in Eq.(\ref{eq:anti-Stokes matrix}), the Rabi period

is inverse proportional to $g$. Therefore the strong coupling regime here can be interpreted as a longer phonon lifetime than the Rabi period. In reported waveguide systems, when pump pulse power is $1W$, a coupling ratio as high as ${g}/{\Gamma} = 8.3$ can be achieved~\cite{Xie:19}. Moreover, due to the decrease of phonon dissipation at low temperatures in optical fibers~\cite{LEFLOCH2003395}, this coupling ratio can even be larger.
\subsubsection{Area Dependency}
SBS can be used to coherently transfer information from the optical domain to acoustic waves. This concept has been experimentally shown as Brillouin memory \cite{zhu2007stored,merklein2017chip}.
In Brillouin memory, the write/read efficiency attains a maximum when the effective coupling area $\Theta(T)=\int_0^T \text{d}t\ g_0 \sqrt{|a_p^\dagger(t)a_p(t)|}$ satisfis the following area dependency equation~\cite{zhu2007stored,Dong:15,dodin2002storing,merklein2017chip}:
\begin{equation}
    \Theta = (2n+1)\frac{\sqrt{2}\pi}{2}\qquad (n\in\mathbb{Z}{})\ .
    \label{eq:AreaDependency}
\end{equation}
This result can be recovered by our formalism in a straightforward way. The readout process in Brillouin memory experiment is the anti-Stokes process, which is described by Eq.~(\ref{eq:anti-Stokes matrix}). We consider the perfect phase-matching case where $\Delta=0$ and omitting the dissipation $\Gamma,\gamma=0$ as in Ref.~\cite{Dong:15}. If the system is driven by a rectangular pump pulse as described in Eq.~(\ref{eq:RectanglePulse}), the matrix elements of the propagator in Eq.~(\ref{eq:propagator_G}) that describes photon-phonon transfer reads:
\begin{equation}
\begin{aligned}
    \frac{1}{2}|\mathbf{G}_{21}(\eta>T)|&=|\mathbf{G}_{12}(\eta>T)|\\
    &=\left|\sin\left(\frac{\sqrt{2}}{2}g T\right)\right|\\
    &=\left|\sin\left(\frac{\sqrt{2}}{2}\Theta(T)\right)\right|\ .
\end{aligned}
\end{equation}
The $|\mathbf{G}_{21}|$ and $|\mathbf{G}_{12}|$ attains maxima if and only if the area dependency in Eq.~(\ref{eq:AreaDependency}) is satisfied.

\subsubsection{The undepleted condition}
\label{sect:undepleted_conditions}
Our formalism is built based on the undepleted pump approximation. When the pump power is strong enough and the pulse length is sufficiently long, the pump power might be significantly depleted by the Stokes process. The anti-Stokes process is much weaker than the Stokes process, and therefore it is enough for us to only consider the Stokes process.

For a short pulse length in the Stokes process, the second term in the solution of Langevin equation Eq.~(\ref{Eq:solution_of_langevin}) can be omitted, {since the first term in Stokes process described by Eq.~(\ref{Eq:Stokes_matrix}) has an exponential growing term, which significantly surpasses the second term.~\cite{Keaton:14}.} From a physical point of view, the Stokes process is a stimulated amplification process. The amplification of the initial state fluctuation is going to be much greater than the additional noise added during the evolutionary process.:
\begin{equation}
\begin{aligned}
        \langle \tilde{a}_{S}^\dagger({\Delta},\eta) \tilde{a}_{S}({\Delta},\eta) \rangle \approx |\mathbf{G}_{12}({\Delta},\eta)|^2 n_{th}\ .
\end{aligned}
\end{equation}
The $n_{th}$ is the average thermal phonon number: $n_{th}\approx{k_B T}/(\hbar \Omega)$. The undeplected condition requires that the Stokes power is much smaller than the pump power: $I_{S}\ll I_P$. For a rectangular pulse with pulse length $T$ as defined in Eq.~(\ref{eq:RectanglePulse}), the requirement can then be simplified to:
\begin{equation}
    \frac{\sqrt{2}}{2}g T \ll \frac{1}{4}\ln \frac{32\pi^2I_p\Omega^2}{G\Gamma c_g k_B^2 T^2 \omega^2}\ .
    \label{eq:undepleted_result}
\end{equation}
The detailed derivation can be found in the appendix. The left-hand side of Eq.(\ref{eq:undepleted_result}) refers to the Rabi area, which equals to $2\pi$ for a complete Rabi period. As an example, for a chalcogenide waveguide~\cite{Xie:19} pumped by $1W$ at room temperature, the right-hand side of Eq.(\ref{eq:undepleted_result}) is $10.27>2\pi$, which means that for those pulses within the first Rabi period, the undepleted condition holds. This result also implies that increasing the pump power while decreasing the pulse length makes the undepleted assumption more robust while keeping the effective coupling area unchanged. The significant increase of depletion threshold, {which is also the Brillouin threshold}, is also predicted and verified experimentally in Ref.~\cite{Keaton:14}.

\section{Applications of Coherent Control}
\label{Section:Application}
In this section, we will use the techniques developed in the previous section to discuss coherent transfer, cooling and entangled pair generation in backward Brillouin scattering. For simplicity, we only consider the case where the waveform of the pump light is a rectangular wave, although the method we proposed earlier is not only applicable to rectangular waves.

\subsection{Coherent Transfer}
\label{sect:coherent_transfer}
The most critical task of optomechanical systems is the manipulation of phonon states. How to store information into phonons, read out the phonon states and convert them into measurable physical quantities therefore becomes an important issue.

\begin{figure}
\centering
\includegraphics[width=0.4\textwidth]{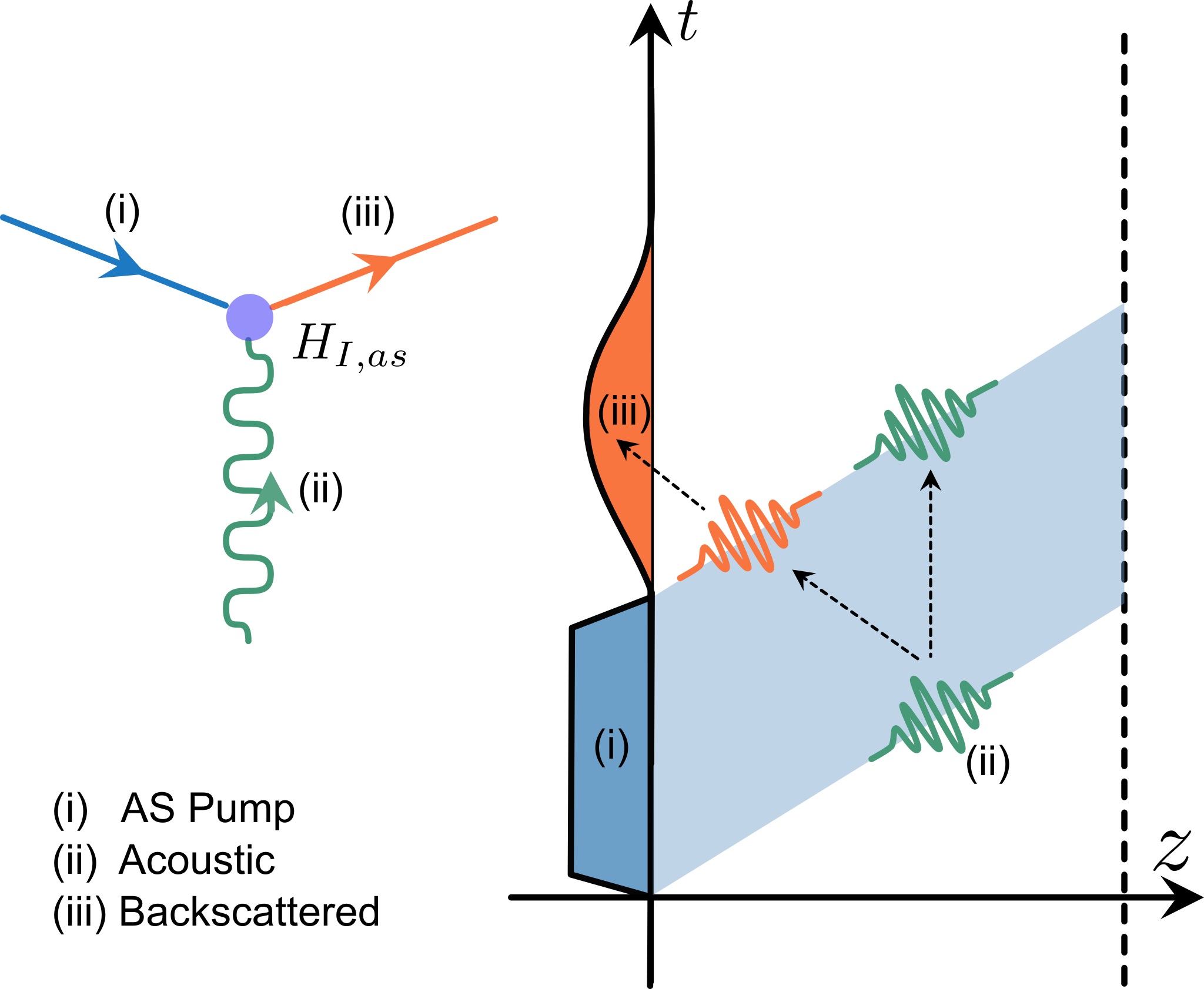}
\caption{Illustration of the coherent transfer and cooling process. The anti-Stokes interaction $H_{I,AS}$ describes a Rabi-oscillation between the anti-Stokes photons and phonons. Thus a $\pi/2$ Rabi pulse can be used to transfer the phonon into photons coherently (vice versa), and once the phonons are transferred into photons, cooling is attained.}

\label{fig:illa_4}
\label{fig:illa_3}
\end{figure}

The main idea to implement phonon readout is to use the anti-Stokes process: the beam-splitter-like Hamiltonian of the anti-Stokes process describes a Rabi oscillation between photons and phonons. Thus, coherent transfer between photons and phonons is possible by controlling the Rabi oscillation. The coherent transfer process can be illustrated using the space-time diagram in FIG.~\ref{fig:illa_4}. The classical Brillouin coherent readout has been demonstrated experimentally on photonic chips~\cite{merklein2017chip}.  As shown in the previous sections, due to the lack of resonating structures in optomechanical waveguides, we must consider the entire continuous phonon spectrum to derive the spectrum dependent coherent-transfer efficiency.

The spectrum dependent read out efficiency $\beta({\Delta},\eta)$ is defined as the following: Assuming that the phonon number at wave vector ${\Delta}$ at the begining of the readout process $t=0$ is: $N_{b}({\Delta},\eta)=\langle \tilde{b}_{as}^\dagger({\Delta,0}) \tilde{b}_{as}({\Delta,0}) \rangle $. After applying the coherent readout process, the phonons will be converted into photons, because of the exisitence of thermal noise, such readout process is only partially coherent: $N_A({\Delta},\eta) = N_{a,c}({\Delta},\eta) + N_{a,n}({\Delta},\eta)$, where the $N_{a,c}({\Delta},\eta) = \langle \tilde{a}_{as}(\Delta,\eta)^\dagger\rangle \langle \tilde{a}_{as}(\Delta,\eta) \rangle$ is the coherent part and $N_{a,n}({\Delta},\eta) = \langle \tilde{a}_{as}(\Delta,\eta)^\dagger \tilde{a}_{as}(\Delta,\eta) \rangle - N_{a,c}({\Delta},\eta)$ is the noise part. Therefore the readout efficiency can be quantified by the following quantity:
\begin{equation}
    \beta({\Delta},\eta) = \frac{N_{a,c}({\Delta},\eta)^2}{\left(N_{a,c}\left({\Delta},\eta\right)+N_{a,n}\left({\Delta},\eta\right)\right)N_{b}\left({\Delta},0\right)}\ .
\end{equation}

\begin{figure}
\centering
\includegraphics[width=0.45\textwidth]{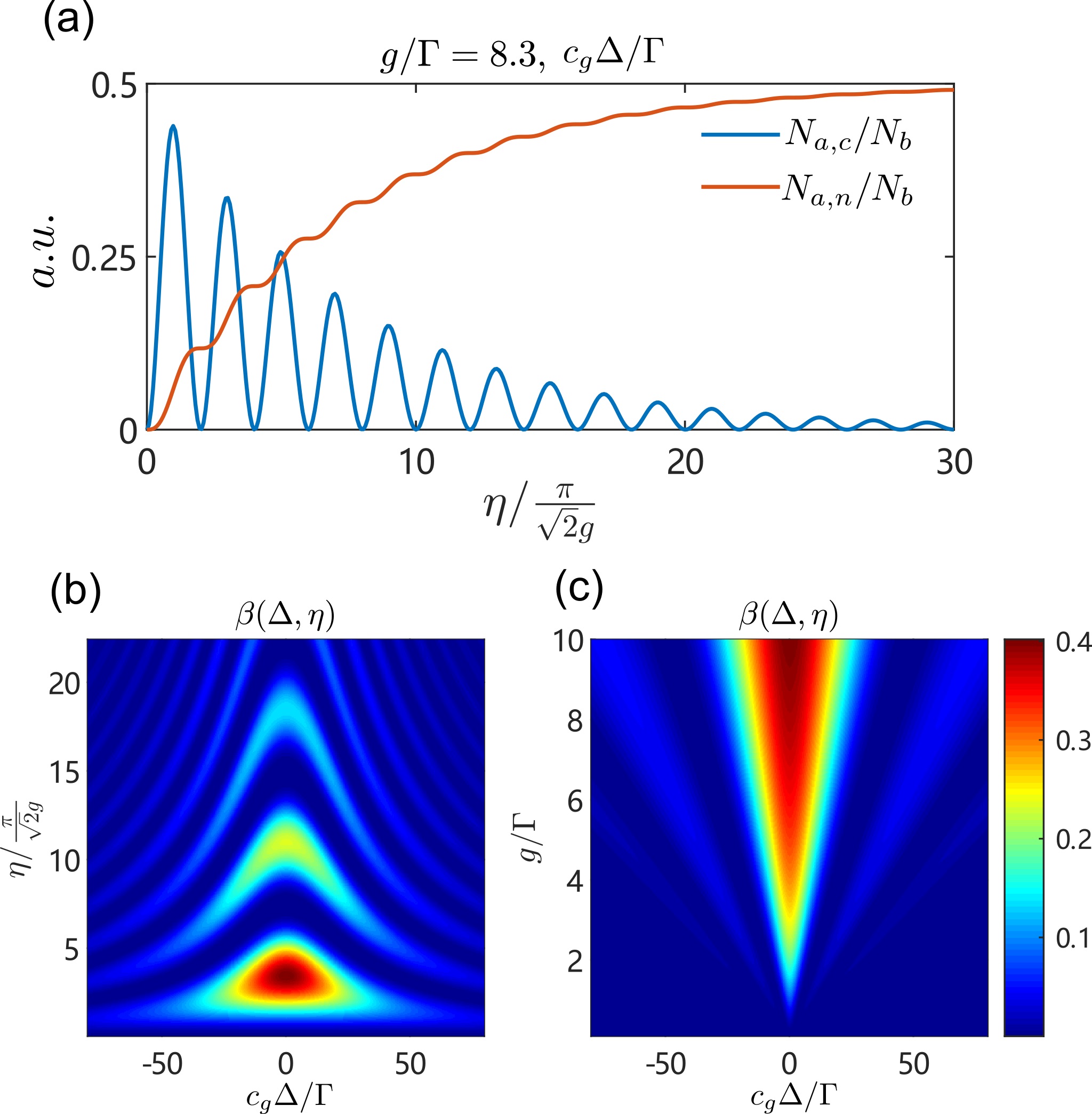}
\caption{The Brillouin coherent transfer. (a)As the pump length increases, the coherent part oscillates, and the body of the phonon is dominated by the noise part when the pulse length is long. (b)The transfer efficiency was calculated using $g=8.3\Gamma$ (Chalcogenide chip~\cite{Xie:19} with pump power equals 1\text{W}). A clear oscillating behavior due to the Rabi nature can be seen. (c) The transfer efficiency $\beta$ at different coupling strengths, when pump length equals the $\pi/2$ Rabi pulse $t=\pi/\sqrt{2}g$. As the coupling increases, the effective transfer spectrum bandwidth will increase. }
\label{fig:transfer_intuitive}
\end{figure}

Figure.\ref{fig:transfer_intuitive}(a) presents how the coherent part $N_{a,c}$ and the noisy part $N_{a,n}$ changes as pulse length increases. Due to the Rabi nature of the anti-Stokes process, the coherent part oscillates while the maximum is attained at the first peak due to the dissipation effect. However, the incoherent noise term accumulates and dominates later. The result shows that effective, coherent transfer can only be possible when the $\pi/2$ Rabi pulse is shorter compared to the phonon lifetime.

Solving the Langevin equation using propagator matrix $\mathbf{G}$ in Eq.(\ref{eq:propagator_G}), the readout photon is:
\begin{equation}
    \begin{aligned}
\tilde{a}_{as}({\Delta}, \eta)=& \mathbf{G}_{11}({\Delta}, \eta) \tilde{a}_{as}({\Delta}, 0)+\mathbf{G}_{12}({\Delta}, \eta) \tilde{b}_{as}({\Delta}, 0) \\
&+\int_{0}^{t} \mathrm{~d} \nu \mathbf{G}_{12}({\Delta}, \eta-\nu) \sqrt{\Gamma n_{t h} / 2} \tilde{\xi}({\Delta}, \nu)\ .
\end{aligned}
\end{equation}

Since we consider that there is no anti-Stokes light in the waveguide at the $t=0$ moment in the readout process, therefore $\tilde{a}_{as}({\Delta},0) = 0$. The spectrum dependent readout efficiency $\beta({\Delta},\eta)$ for a rectangular pump pulse with length $t$ reads:
\begin{equation}
    \beta({\Delta},\eta) = \frac{|\mathbf{G}_{12}({\Delta},\eta)|^4}{|\mathbf{G}_{12}({\Delta},\eta)|^2+|\Gamma \int_{0}^\eta \text{d}\nu\ \mathbf{G}_{12}({\Delta},\nu)|^2}\ ,
\end{equation}
where
\begin{equation}
            \mathbf{G}_{12} = -\mathrm{i}\mathrm{e}^{-\frac{1}{4}\Gamma_{e, as}^* \eta-\mathrm{i}c_g\Delta \eta}\frac{g}{\sqrt{2}g_{e, as}}  \sin \left( \frac{\sqrt{2}}{2}g_{e, as} \eta\ \right)\ .
\end{equation}
The resonance modified effective coupling strength $g_e$ and the effective acoustic dissipation are defined as:
\begin{equation}
    \begin{aligned}
g_{e,as} &=\sqrt{g^{2}-(\Gamma+\mathrm{i}c_g\Delta)^{2}/8}\ , \\
\Gamma_{e,as} &=\Gamma+ \mathrm{i} c_g {\Delta}\ .
\end{aligned}
\label{eq:new_variable}
\end{equation}
The detailed derivation of $\mathbf{G}_{12}$ can be found in Appendix.\ref{appendix:cooling}.

We present the numeric result for $\beta({\Delta},\eta)$ in Fig.\ref{fig:transfer_intuitive}(b) and (c). As shown in Fig.\ref{fig:transfer_intuitive}(b), the readout efficiency oscillates as the pulse length increases, which agrees with the area dependency law in Brillouin memory \cite{dodin2002storing}. The highest coherent readout (highest $\beta$) can be achieved when:
\begin{equation}
\eta\approx \frac{\pi}{\sqrt{2}g}
    \label{eq:coherent_eff}\ .
\end{equation}
This is exactly the $\pi/2$ Rabi pulse length slightly modified due to the dissipation effect. Figure.\ref{fig:transfer_intuitive}(c) shows the readout efficiency $\beta$ calculated at $\eta= {\pi}/({\sqrt{2}g})$, which is the $\pi/2$ Rabi pulse. It can be seen that both larger readout bandwidth and higher maximum readout efficiency can be achieved by increasing the coupling strength $g$. Both Fig.\ref{fig:transfer_intuitive}(a) and (b) show oscillating behaviors along the wave vector ${\Delta}$, which is related to the higher harmonics components in the Fourier transformation of rectangular waves.

\subsection{Brillouin Cooling}

Many quantum experiments can only be carried out at low temperatures (passive cooling) in order to reduce the decoherence effect introduced by thermal noise. Laser cooling is one of the most promising active cooling techniques \cite{RevModPhys.86.1391} to reduce thermal noise in a specific frequency band.
Recently, laser cooling induced by anti-Stokes Brillouin scattering, i.e. Brillouin cooling, has been explored in optomechanical waveguides~\cite{ PhysRevX.8.041034,chen2016brillouin} by utilizing forward Brillouin scattering where phonons experience lower damping than photons. However, Brillouin cooling generated by backward Brillouin scattering where the acoustic dissipation exceeds the optical dissipation in typical Brillouin-active waveguides is still largely unexplored. In this section, we proposed a cooling mechanism by using pulses rather than continuous waves. We claim that this cooling mechanism might attain higher cooling efficiency and can be used in long fibers.

Conventional laser cooling is based on increasing the effective dissipation of phonons by using a damping laser. Here we propose an alternative cooling mechanism based on coherent transfer. As shown in the previous section, the anti-Stokes process can be regarded as a photon-phonon Rabi oscillation using the linearized effective Hamiltonian. The main idea of coherent transfer-based cooling is to use a carefully designed laser pulse to convert phonons to photons through the phonon-photon Rabi oscillations.  Those photons that are converted from phonons will leave the waveguide since the waveguide is open, which is different from cavities. Our laser pulses at the appropriate pulse length will avoid reverse conversion, thus leaving fewer phonons in the waveguide system. By using a pulsed pump, the pump depletion effect due to the Stokes process can also be avoided. The cooling process is illustrated using space-time diagram in FIG.~\ref{fig:illa_3}.

The cooling effect can be quantified by counting the remained phonons. We introduce the spectrum remained phonon rates $\kappa({\Delta},\eta)$:
\begin{equation}
    \begin{aligned}
        \kappa({\Delta},\eta) &= \frac{\langle \tilde{b}^\dagger({\Delta},\eta) \tilde{b}({\Delta},\eta)\rangle }{\langle \tilde{b}^\dagger({\Delta},0) \tilde{b}({\Delta},0) \rangle} \\
        &= \left|\mathbf{G}_{22}({\Delta}, \eta)\right|^2  +\Gamma \int_{0}^{\eta}\text{d}\nu\  \left|\mathbf{G}_{22}({\Delta}, \nu)\right|^2 \\
        &= \kappa_{c} + \kappa_{n}\ .
    \end{aligned}
    \label{eq:cooling_intuitive}
\end{equation}
The smaller the $\kappa({\Delta},\eta)$, the better cooling effect is achieved.

\begin{figure}
\centering
\includegraphics[width=0.45\textwidth]{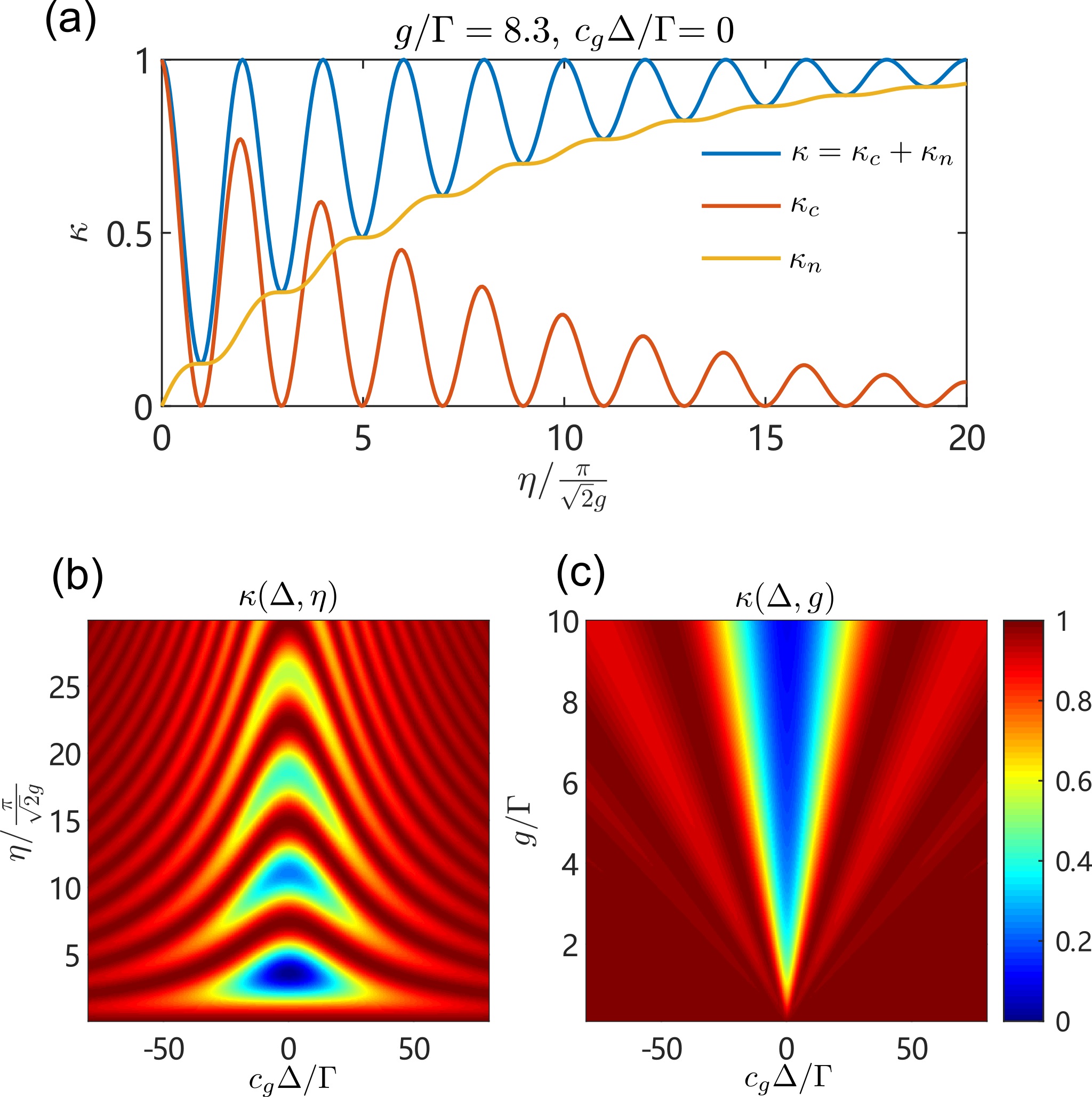}
\caption{The Brillouin cooling. (a)The remained phonon consists of two parts: the coherent part determined by the initial state and the noise generated part. Optimal cooling can be achieved by choosing propriate pump length to be $\pi/2$ Rabi period. (b)The transfer efficiency was calculated using $g=8.3\Gamma$ (Chalcogenide chip~\cite{Xie:19} with pump power equals 1W). A clear oscillating behavior due to the Rabi nature can be seen. (c)The cooling effect on different parameters, when pump length equals the $\pi/2$ Rabi pulse $\eta=\pi/\sqrt{2}g$.. As the coupling strength increases, the effective cooling bandwidth will increase.}
\label{fig:cooling_intuitive}
\end{figure}

Eq.(\ref{eq:cooling_intuitive}) consists of two terms, the first one $\kappa_c$ referring to the coherent transfer of the initial phonon and the second one $\kappa_n$ to the increased thermal noise. Fig.\ref{fig:cooling_intuitive} shows the two terms and the whole $\kappa({\Delta}=0,t)$ at the perfect phase-matching point ${\Delta}=0$. One can clearly see the Rabi oscillation behavior of the coherent part, which contributes to cooling, and the accumulating thermal noise counterbalances the cooling effect, which eliminates the cooling effect when the pulse length increases.

In conventional fiber optic systems, the dissipation rate of the optical channel is much smaller than the dissipation rate of the acoustic channel: $\gamma \ll \Gamma$. With this approximation, spectrum remained phonon rate reads:
\begin{equation}
    \kappa({\Delta}, \eta)\approx 1-\left|\mathrm{e}^{-\frac{\Gamma_{e,as}\eta}{2}}\frac{8 g_{e, as}^{2} + \Gamma_{e, as}^{2}}{8 g_{e, as}^{2}}
    \, \sin ^{2}\left(\frac{g_{e, as} \eta}{\sqrt{2}}\right)\right|\ .
    \label{eq:cooling_approx_result}
\end{equation}

The $g_{e, as}$ and $\Gamma_{e,as}$ is defined in Eq.(\ref{eq:new_variable}). The detailed derivation for the above result can be found in the Appendix.\ref{appendix:cooling}. Eq.(\ref{eq:cooling_approx_result}) shows that, choosing
\begin{equation}
    \eta=(2n+1)\frac{\pi}{\sqrt{2}g}\qquad (n\in\mathbb{Z})
    \label{eq:Rabi_cooling}\ ,
\end{equation}
leads to the minimum $\kappa({\Delta},\eta)$, which is equivalent to the maximum cooling efficiency. Due to the accumulating thermal noise, minimum $\kappa({\Delta}=0,\eta)$ can be achieved near the first Rabi $\pi/2$ pulse at $\eta\approx \pi/(\sqrt{2}g)$, this is the same as the maximum coherent transfer efficiency shown in Eq.(\ref{eq:coherent_eff}), which is also based on the same photon-phonon Rabi oscillation.

\begin{figure}
\centering
\includegraphics[width=0.35\textwidth]{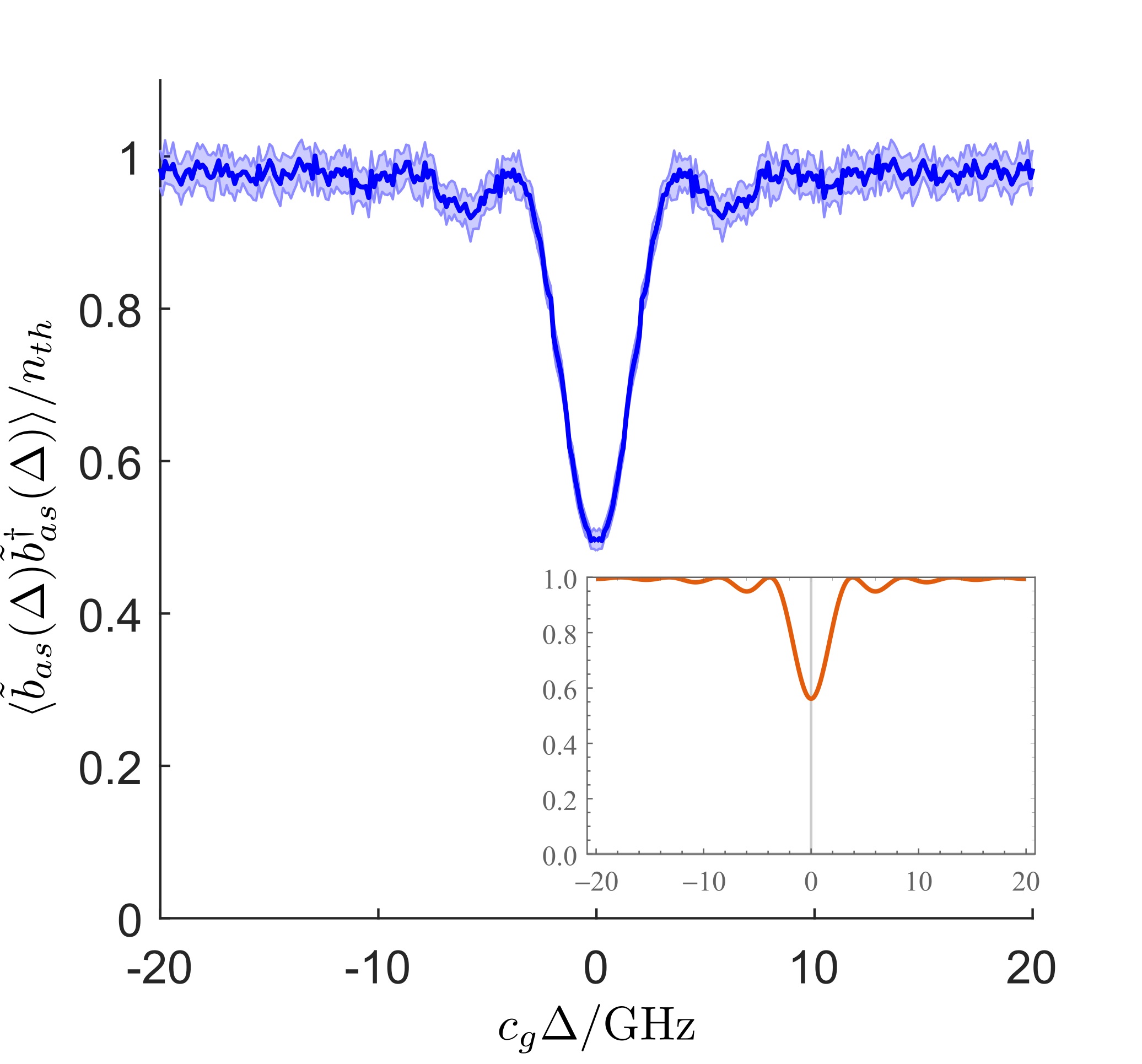}
\caption{The remained phonon spectrum after a $\pi/2$ cooling pulse. This figure is obtained by the first-principle simulation using the algorithm from Ref.~\cite{Nieves:21,deSterke:91}. In the simulation, a chalcogenide waveguide~\cite{Xie:19} with length $L=1\text{m}$ is assumed. A $\pi/2$ Rabi pulse with pump power $I_p = 10\text{W}$, pulse length $t=\frac{\pi}{\sqrt{2}g}$ is used as the cooling pulse. The phonon spectrum is calculated when the pump pulse leaves the fiber at $t = L/c_g$. The shaded area indicates the $3\sigma$ error, and the theoretically predicted result is shown in the lower right corner. By applying a rectangular cooling pulse, the remaining phonon spectrum is sinc-like, and around $50\%$ of the initial phonon number at the central wave vector has been removed. }
\label{fig:cooling_spectrum}
\end{figure}

In Fig.\ref{fig:cooling_intuitive}(b) and (c), we present the calculated $\kappa({\Delta},\eta)$ for different coupling strength and pulse length by assuming rectangular pump pulse. Due to the decoherence effect of thermal noise, the optimal cooling pulse is the $\frac{\pi}{2}$ Rabi-pulse, which refers to $n=1$ case in Eq.(\ref{eq:Rabi_cooling}). From Fig.\ref{fig:cooling_intuitive}(c), one can clearly see that the cooling bandwidth becomes wider for stronger coupling, which indicates a wider interaction bandwidth. We claim that this has the same mathematical roots as the general power broadening effect in all atomic systems~\cite{citron1977experimental}.

\subsection{Entanglement}

Quantum entanglement is one of the fundamental building blocks of today's quantum technologies, especially quantum communication. The generation of entangled quantum pairs is the basis of quantum state teleportation and quantum repeaters. Classical information networks based on optical fibers are one of the most promising infrastructures for future quantum teleportation, which means that generating entangled pairs in an all-fiber system is a fruitful challenge. Apart from the application-based perspective, achieving the generation of entangled pairs is also one of the vital experiments to demonstrate the ability to do quantum experiments in optomechanical waveguides.

\begin{figure}
\centering
\includegraphics[width=0.40\textwidth]{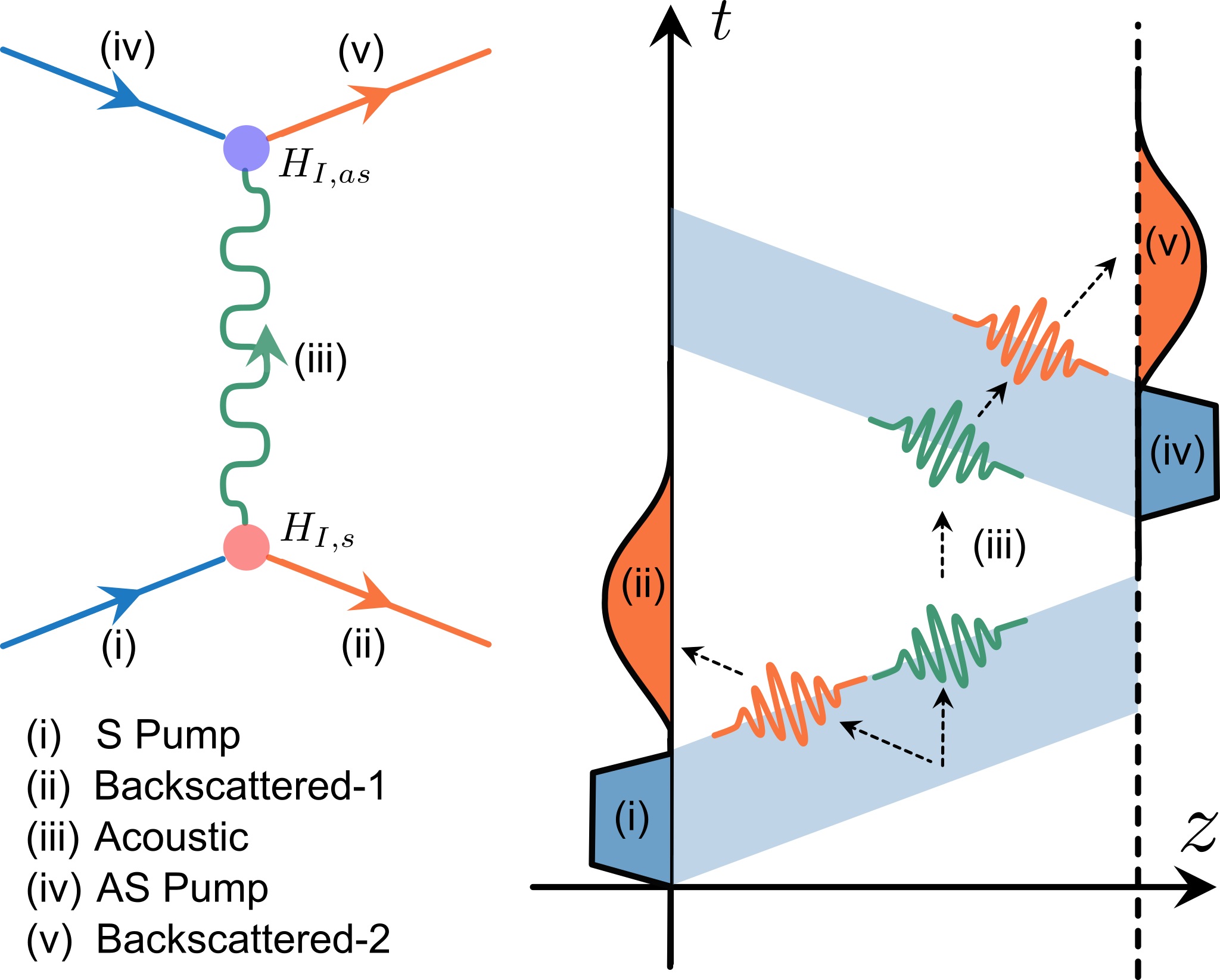}
\caption{Illustration of the entangled pair generation process. First, a blue-detuned pump pulse (S pump) should be applied to generate the entangled phonon-photon pairs via the Stokes interaction $H_{I, s}$. Then, a red-detuned pump pulse (AS pump) should be applied to stimulate the coherent transfer process via anti-Stokes interaction $H_{I, as}$ described in the previous sections to readout the phonon into a photon. By combining these two processes, the measurable entangled photons pairs (Backscattered-1 and Backscattered-2) could be generated.}
\label{fig:illa_5}
\end{figure}

In this section, we show that the photon-phonon entangled pair generation can be achieved by utilizing the Stokes process, and the entangled photon-phonon pairs can be further transformed into photon-photon entangled pairs by the coherent transfer technique described in Sec.~\ref{sect:coherent_transfer}, as shown in terms of the space-time diagram in FIG.~\ref{fig:illa_5}.  Choosing the Stokes process to generate photon-phonon pairs is motivated by the down-conversion nature of the Stokes process. In the Stokes process, a higher energy photon is annihilated, producing a lower energy phonon and a lower frequency photon. In this process, both momentum and energy conservation must be satisfied, which leads to the phase-matching condition. As a direct result of this phase-matching condition, there is some shared information between the produced phonon and photon, which leads to quantum entanglement.

We introduce the quadrature operators for the acoustic field and the optical field:
\begin{equation}
    \begin{aligned}
X_{a} &=\frac{a+a^{\dagger}}{\sqrt{2}}\ , \quad Y_{a} =\mathrm{i} \frac{a^{\dagger}-a}{\sqrt{2}} \ ,\\
X_{b} &=\frac{b+b^{\dagger}}{\sqrt{2}}\ , \quad Y_{b} =\mathrm{i} \frac{b^{\dagger}-b}{\sqrt{2}} \ .
\end{aligned}
\end{equation}

To prove the existence of quantum entanglement, we use Duan's two-mode entanglement criterion~\cite{PhysRevLett.84.2722}. The main idea of the Duan's criterion is to choose two EPR variables and calculate the variance sum. It can be shown that once the correlation variance is less than a specific quantum limit, the density matrix of the two quantum modes cannot be separated by any means, thus leading to quantum entanglement.

\begin{figure}
\centering
\includegraphics[width=0.5\textwidth]{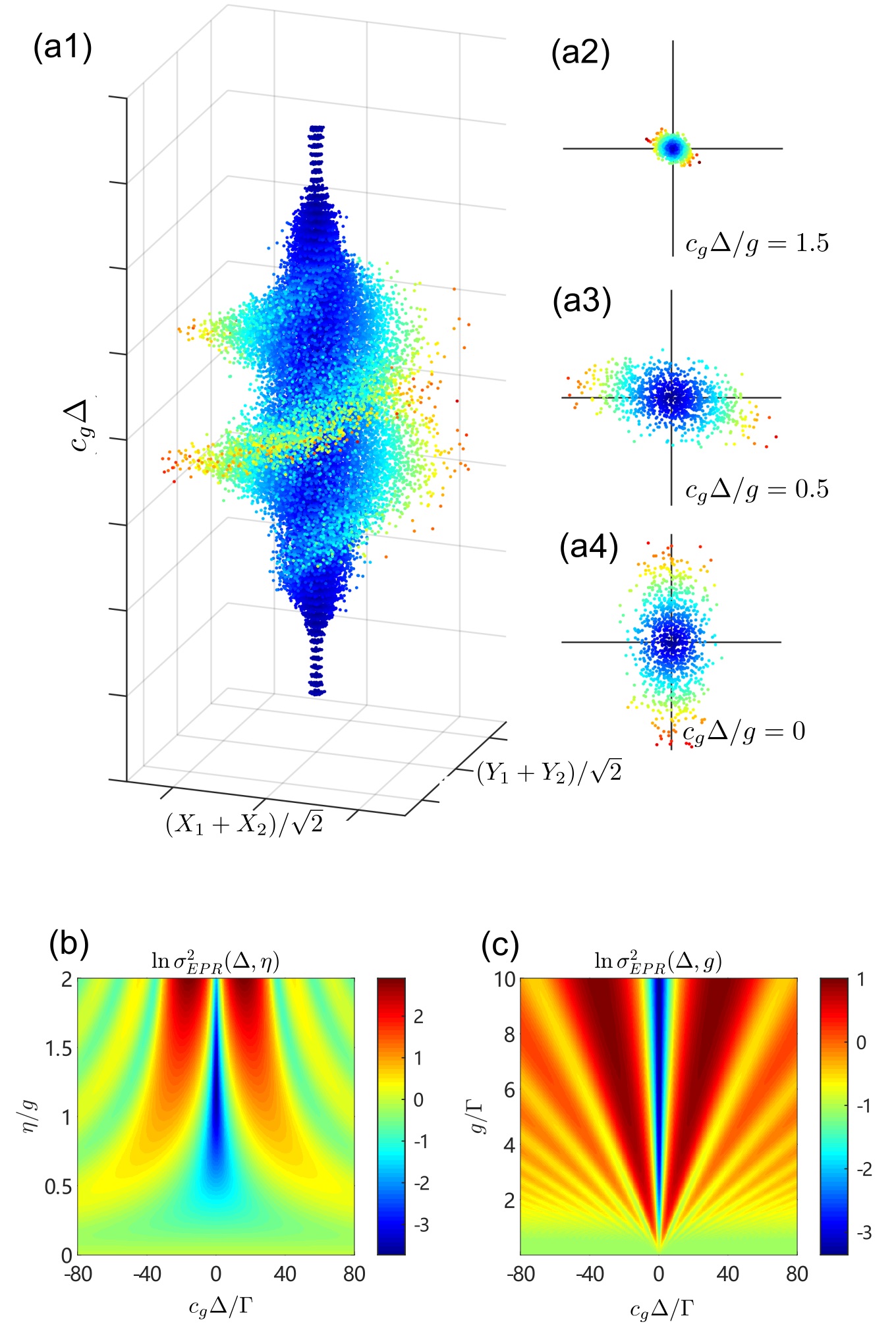}
\caption{The Brillouin entanglement. (a1)The quadrature distribution in different ${\Delta}$ is shown using algorithm from Ref.~\cite{Nieves:21,deSterke:91}). The simulation parameter is the same as Fig.\ref{fig:cooling_spectrum}. $X_1,X_2,Y_1,Y_2$ are the quadrature components of the two backscattered pulses. Non-uniform distribution of quadratures can be seen when $|{\Delta}|<\Gamma$, which implies shared information between two scattered photons. When $|{\Delta}|$ is large, the Brillouin interaction strength is limited by the phase-matching condition, so the two-mode correlation between quadratures gradually disappears. The discretization along the frequency axis ($c_g\Delta$) is the result of the Fourier transform of the finite numerical simulation time. a(2-4) is the transection of (a1) at the three points of wavevector $|{\Delta}|$ from larger to smaller. (b,c)The EPR variance $\sigma^2_{EPR}$. As the coupling increases, the effective entangled spectrum bandwidth will increases. (b) is calculated using $g=8.3\Gamma$ (Chalcogenide chip~\cite{Xie:19} with pump power equals 1W). }
\label{fig:entangle_phonon_photon}
\end{figure}

The two mode EPR variables we choosed are:
\begin{equation}
    \begin{gathered}
u=\frac{1}{\alpha} X_{a}+ \alpha Y_{b} \ ,\\
v=\frac{1}{\alpha} Y_{a}+\alpha X_{b} \ ,
\end{gathered}
\end{equation}
with $\alpha = 2^{-1/4}$. {In the definition of EPR variance, $\alpha$ can be freely chosen~\cite{PhysRevLett.84.2722}. This particular $\alpha$ value is obtained by trying to minimize the EPR variance.} The motivation of such special choice of $\alpha$ comes from the unbalanced interaction time of optical channel and acoustic channel due to the backscatter nature, which is also described by the unbalanced interaction term $g/2$ and $g$ in Eq.(\ref{Eq:Stokes_matrix}). The Duan's entanglement criterion for it is:
\begin{equation}
        \sigma^2_{EPR}=\frac{\sigma^2 u+\sigma^2 v }{\alpha^2 + \alpha^{-2}}< 1\ .
\end{equation}
After some calculations, the expression of $\sigma^2_{EPR}$ reads:
\begin{equation}
    \begin{aligned}
             \sigma^2_{EPR}= & \frac{1}{3}\left|\sqrt{2}\mathbf{G}_{11}+\mathrm{i}\mathbf{G}_{21}\right|^2
             +\frac{2n_{0}+1}{3} \left|\sqrt{2}\mathbf{G}_{12}
             +\mathrm{i}\mathbf{G}_{22}\right|^2
             \\
             & \quad
             +\frac{2n_{th}+1}{3} \Gamma \int_0^\eta \text{d}\nu \left|\sqrt{2}\mathbf{G}_{12}+\mathrm{i}\mathbf{G}_{22}\right|^2
             \\
             &\approx \left(1 + \frac{2n_0}{3}\right)|G(\eta)|^2 +
             \frac{2n_{th}+1}{3}\Gamma\int_0^\eta d\nu\  |G(\nu)|^2\ ,
    \end{aligned}
\end{equation}
where
\begin{equation}
G(\eta) = \mathrm{e}^{-\frac{1}{4} \Gamma_{e,s}^* \eta}\left[ \mathrm{e}^{-\frac{g_{e,s} \eta}{\sqrt{2}}}-  \frac{\Gamma_{e,s}}{2\sqrt{2}g_{e,s}}\sinh \left( \frac{ g_{e,s} \eta }{\sqrt{2}}\right) \right]\ .
\end{equation}
The detailed calculation can be found in Appendix.\ref{appendix:entangle}. The $n_0$ is the phonon number expectation at the initial state $\eta=0$, and the $n_{th}$ is the thermal phonon expecation number determined by the temperature of the enviornment. When the system is cooled by the cooling technique we proposed in the previous sections, $n_0<n_{th}$ can be achieved. The off-resonance effective coupling and the effective dissipation for Stokes process shown here are slightly different from what we defined in the anti-Stokes process in Eq.(\ref{eq:new_variable}):
\begin{equation}
    \begin{aligned}
        g_{e,s} &= \sqrt{g^2 + (\Gamma+\mathrm{i}c_g\Delta)^2/8}\ ,\\
        \Gamma_{e,s} &= \Gamma + \mathrm{i}c_g{\Delta}\ .
    \end{aligned}
\end{equation}

The $G(\eta)$ consists of two competing terms. The first term shows an exponential depressing of the EPR variance with rate $g_{e,s}$ while the sinh-like second term will refer to the decoherence effect that will destroy such entanglement. Therefore the photon-phonon entangle can be achieved by choosing adequate pump length in the Stokes process.

We present the numerical result for the photon-phonon entangled EPR variance in Fig.\ref{fig:entangle_phonon_photon}. The dashed lines in Fig.\ref{fig:entangle_phonon_photon}(b) and (c) refer to the quantum non-separation limit $\sigma^2_{EPR}=1$. The generation of photon-phonon entanglement pairs is based on a down-conversion-like Stokes process rather than a beam-splitter-like anti-Stokes process. Therefore, unlike the coherent transport and cooling discussed in the previous sections, there is no Rabi oscillation behavior. As shown in the Fig.\ref{fig:entangle_phonon_photon}(a), the longer pulse will induce a stronger entangled effect while decreasing the entangled bandwidth, and stronger coupling can still broaden the entangled bandwidth. In practice, the optimal entangled pulse length should be specified by taking the measurement bandwidth of the experiment setup into consideration: the photon-phonon entangled pair cannot be measure directly, the coherent transfer process is needed to transfer the entangled phonon into a photon, in which the measurement bandwidth can be controlled by manipulating the pump waveform as discussed in previous sections. We show in Fig.\ref{fig:entangle_phonon_photon} the results of photon-phonon entangled pair generation obtained by direct computation with the noisy Brillouin simulation algorithm~\cite{Nieves:21}. The simulation combines photon-phonon entangled pair generation with the coherent transfer, which is directly computed for photon-phonon entangled pairs.

\section{Conclusion and Perspectives}
\label{Conclusion}
In this paper, we proposed a formulation to treat traveling phonons and traveling photons in a optomechanical waveguide system as an array of multiple optomechanical cavities. The possibility of realizing coherent control using this formulation has been shown: Both the effective coupling strength and the interaction time can be controlled easily by manipulating the shape of the pump pulse. By applying the formulation, we show that it is possible to achieve active Brillouin cooling through the backward Brillouin scattering process and quantum entangled pair generation in waveguide systems. Experiments based on this formalism are shown to be achievable using current technology and fabrication of optical fibers and integrated waveguides.

Our work mainly focuses on the backward scattering optomechanical interaction and most of the calculations are done using the assumption that the pump pulse is almost non-depleted. The undepleted assumption leads to a linearized Hamiltonian directly, which is similar to down-conversion for the Stokes process and similar to a beam-splitter for the anti-Stokes process. The available quantum operation time approximately equals the ratio between the coupling strength and the dissipation rate: $N_{op} = g/\Gamma$. There are two routes, including the dissipation reduction and coupling enhancement, towards a broader range of quantum applications within optomechanical waveguides. In order to reduce the dissipation rate, efforts have to be paid in designing waveguides that can localize and trop acoustic phonons with high efficiency. In order to increase the coupling strength, one method is to increase the material-dependent coupling strength $g_0$ by optimizing the waveguide structure or using more promising materials~\cite{Hatanaka2014}. Another method to increase the coupling strength is to use higher pump powers. In this case, the pump light may be significantly depleted, so shorter light pulses are needed to meet the requirements of coherent control. Therefore, the undepleted regime within the short pulses regime (below 100\,ps) need to be considered in future works.

Because all Brillouin backscattering processes under undepleted conditions can be solved exactly based on the method of this paper, future efforts should be spent on detailed modeling of the Brillouin memory process, where information is coherently stored in acoustic waves \cite{zhu2007stored,merklein2017chip}. In particular, the method can be the base for a modell including quantum noise and a study on how coherent information is destroyed by different noise contributions.

\section*{Acknowledgements} \label{sec:acknowledgements}
    {Authors acknowledge funding from the Max Planck Gesellschaft through an independent Max Planck Research Group, CW acknowledges funding from the Villum Fonden related to grant No.~16498.}
\bibliography{bibliography}

\appendix*
\onecolumngrid
\newpage
\appendix

\subsection{Detailed Derivation of the Undepleted Conditions}
\label{appendix:undepleted}
The backscattered Stokes power is the integral of photon densities:
\begin{equation}
    I_S = \int_{-\infty}^{+\infty} \text{d}{\Delta}\ \frac{\hbar\omega c_g}{2\pi} \langle \tilde{a}_{s}^\dagger(\Delta) \tilde{a}_{s}(\Delta) \rangle\ .
\end{equation}
Using the technique we developed, we have:
\begin{equation}
    \langle \tilde{a}_s^\dagger(\Delta,\eta)\tilde{a}_s(\Delta,\eta)\rangle  = |\mathbf{G}_{12}({\Delta},\eta)|^2\langle \tilde{b}_s^\dagger(\Delta,0) \tilde{b}_s(\Delta,0)\rangle +n_{th} \Gamma \int_0^t\text{d}\nu |\mathbf{G}_{12}({\Delta},\nu)|^2 \ .
\end{equation}
For the system driven by short pulses $\eta\sim 1/g\ll 1/\Gamma$ in strong coupling regime, the second term which is the noise contribution can be omitted. The average thermal phonon number at $t=0$ reads:
\begin{equation}
    \langle \tilde{b}_s^\dagger(\Delta,0) \tilde{b}_s(\Delta,0)\rangle = \frac{1}{\exp{\frac{\hbar\Omega}{k_B T}-1}} \approx \frac{k_B T}{\hbar\Omega}\ .
\end{equation}
Therefore the spectrum density of Stokes photon reads:
\begin{equation}
    \langle \tilde{a}_s^\dagger(\Delta,\eta)\tilde{a}_s(\Delta,\eta)\rangle\approx \frac{k_B T}{\hbar\Omega}|\mathbf{G}_{12}({\Delta},\eta)|^2\ .
\end{equation}
For $g\ll \Gamma$, the following approximation holds:
\begin{equation}
    \mathbf{G}_{12}({\Delta},\eta)\approx \frac{2g\sinh{\frac{1}{4}\sqrt{8g^2-c^2{\Delta}^2}}t}{\sqrt{8g^2-c^2{\Delta}^2}}\ .
\end{equation}
The peak of $\mathbf{G}_{12}(\Delta,\eta)$ is:
\begin{equation}
\begin{aligned}
            \mathbf{G}_{12}(\Delta=0,\eta) &= \frac{\mathrm{i} \sinh \frac{g\eta}{\sqrt{2}}}{\sqrt{2}}
            \approx \frac{\mathrm{i} \mathrm{e}^{g\eta/\sqrt{2}}}{2\sqrt{2}}\ .
\end{aligned}
\end{equation}
The width of the central peak can be approximated by solving $8g^2-c^2\Delta^2=0$:
\begin{equation}
    \text{Wd.} \approx 2\times\frac{2\sqrt{2}g}{c}\ .
\end{equation}
Finally we have:
\begin{equation}
    \begin{aligned}
            I_S &\approx  \int_{-\infty}^{+\infty} \text{d}{\Delta}\ \frac{\hbar\omega c_g}{2\pi} \frac{k_B T}{\hbar\Omega}|\mathbf{G}_{21}(\Delta,\eta)|^2\\
            &\approx \frac{\hbar\omega c_g}{4\pi} \frac{k_B T}{\hbar\Omega}(\text{Wd.}\times|\mathbf{G}_{12}(\Delta=0,\eta)|^2)
            &=\frac{\mathrm{e}^{\sqrt{2}g\eta}gk_B T\omega}{2\sqrt{2}\pi\Omega}\ .
    \end{aligned}
\end{equation}
The undepleted condition is satisified if and only if the bakscattered Stokes power is much smaller than the pump power: $I_S\ll I_P$. Since the effective coupling $g$ is related to the pump power $I_P$ by:
\begin{equation}
    g = \sqrt{\frac{G I_P \Gamma c_g}{4}}\ ,
\end{equation}
where the $G$ is the Brillouin gain, the above undepleted condition can be simplified into:
\begin{equation}
    g t \ll \frac{1}{2\sqrt{2}}\ln \frac{32\pi^2I_p\Omega^2}{G\Gamma c_g k_B^2 T^2 \omega^2}\ .
\end{equation}

\subsection{Detailed Derivation for Brillouin Cooling}
\label{appendix:cooling}
In this section, we present the detailed derivation process for Brillouin cooling. Because the phonons in the anti-Stokes process and Stokes process are separated by their opposite travel direction, we can only consider the anti-Stokes process, which describes the photon-phonon transfer.

The phenomenologically introduced effective Hamiltonian for anti-Stokes process reads (Eq.(\ref{eq:anti-Stokes matrix})):
\begin{equation}
\begin{aligned}
\frac{\partial}{\partial \eta}
 \begin{pmatrix}
\tilde{a}_{as}(\Delta,\eta) \\ \tilde{b}_{as}(\Delta,\eta)
\end{pmatrix}
=
\begin{pmatrix}
-\mathrm{i} c_g {\Delta}/2-\gamma/4 & - \mathrm{i} g/2\\
- \mathrm{i} g & -\mathrm{i}c_g\Delta-\Gamma/2
\end{pmatrix}
\begin{pmatrix}
\tilde{a}_{as}(\Delta,\eta) \\ \tilde{b}_{as}(\Delta,\eta)
\end{pmatrix}+
\begin{pmatrix}
0\\ \sqrt{\Gamma }\tilde{\xi}
\end{pmatrix}\ .
\end{aligned}
\end{equation}
The $c_g$ is the group velocity of the optical field near the phase matching point. The $g$ is the effective coupling strength enhanced by the pump power, and $n_{th}$ is the averaged thermal phonon number. The relation between the effective coupling and the well known Brillouin coupling strength $G$ is (Eq.(\ref{eq:coupling})):
\begin{equation}
    g = \sqrt{\frac{G P \Gamma c_g}{4}}\ .
\end{equation}
This can be obtained by calculating the steady-state behavior of the coupled mode equation, as shown in the main context of this paper.
The $\tilde{\xi}$ describes the thermal noise, the thermal behavior of phonon determines both the initial state and the noise term, which could be described by a Wigner process:
\begin{equation}
    \begin{aligned}
&\left\langle \tilde{b}_{as}^{\dagger}(\Delta,\eta=0) \tilde{b}_{as}(\Delta,\eta=0)\right\rangle=n_{t h} \ ,\\
&\left\langle\tilde{\xi}^{\dagger}\left(\eta_{1}\right) \tilde{\xi}\left(\eta_{2}\right)\right\rangle=\delta(\eta_1-\eta_2)\ .
\end{aligned}
\end{equation}.

The equation is a Langevin equation. Therefore we can solve the equation using the conventional method for Langevin equations, which is the undetermined coefficient method. Considering the rectangular pump wave, we introduce the $\mathbf{P}$ matrix as:
\begin{equation}
    \mathbf{P}=
    \begin{pmatrix}
-\mathrm{i} c_g {\Delta}/2-\gamma/4 & - \mathrm{i} g/2\\
- \mathrm{i} g & -\mathrm{i}c_g\Delta-\Gamma/2
\end{pmatrix}\ .
\end{equation}
For a Langevin equation in the form:
\begin{equation}
    \frac{d}{d \eta} \myvect{\mathbf{M}}=\mathbf{P} \myvect{\mathbf{M}}+\myvect{\mathbf{R}}\ .
\end{equation}
The solution reads:
\begin{equation}
    \myvect{\mathbf{M}}(\eta)=\exp(\mathbf{P} \eta) \myvect{\mathbf{M}}(0)+\int_{0}^{\eta} \exp[\mathbf{P}(\eta-\nu)] \myvect{\mathbf{R}}(\nu) d \nu\ .
\end{equation}
Therefore we need to calculate the matrix exponential of the $\mathbf{P}$ matrix. The matrix exponential can be calculated by using the formula:
\begin{equation}
    \mathrm{e}^{\mathbf{S}\mathbf{D}\mathbf{S}^{-1}} = \mathbf{S}\mathrm{e}^{\mathbf{D}}\mathbf{S}^{-1}\ .
\end{equation}
Where the Jordan decomposition is used:
\begin{equation}
    \mathbf{P}=\mathbf{S D S}^{-1}\ .
\end{equation}
The $\mathbf{S}$ is the similar matrix, and $\mathbf{D}$ is the Jordan matrix. In our case, $\mathbf{D}$ is diagonalized. By introducing the small optical dissipation approximation:
\begin{equation}
    \begin{aligned}
&(\Gamma \pm \gamma) \approx \Gamma\ . \\
\end{aligned}
\end{equation}
Under those approximations the matrix exponential can be obtained:
\begin{equation}
    \mathbf{G}=\exp(\mathbf{P} \eta)\ .
\end{equation}
The matrix elements reads:
\begin{equation}
    \begin{aligned}
        \mathbf{G}_{11} &= \mathrm{e}^{-\frac{1}{4} \Gamma_{e, as}^* \eta-\mathrm{i}c_g\Delta \eta} \left[ \cos \left(\frac{g_{e, as} \eta}{\sqrt{2}}\right) + \frac{\Gamma_{e, as}}{2\sqrt{2}g_{e, as}}\sin\left(\frac{g_{e, as} \eta}{\sqrt{2}}\right)\right]\ ,\\
        \mathbf{G}_{12} &= -\mathrm{i}\mathrm{e}^{-\frac{1}{4}\Gamma_{e, as}^* \eta-\mathrm{i}c_g\Delta \eta}\frac{g}{\sqrt{2}g_{e, as}}  \sin \left(\frac{g_{e, as} \eta}{\sqrt{2}}\right)\ ,\\
        \mathbf{G}_{21} &= -\mathrm{i}\mathrm{e}^{-\frac{1}{4}\Gamma_{e, as}^* \eta-\mathrm{i}c_g\Delta \eta}\frac{g}{\sqrt{2}g_{e, as}}  \sin \left( \frac{g_{e, as} \eta}{\sqrt{2}} \right) \ ,\\
        \mathbf{G}_{22} &= \mathrm{e}^{-\frac{1}{4} \Gamma_{e, as}^* \eta-\mathrm{i}c_g\Delta \eta} \left[ \cos \left(\frac{g_{e, as} \eta}{\sqrt{2}} \right) -\frac{\Gamma_{e, as}}{2\sqrt{2}g_{e, as}}\sin\left(\frac{g_{e, as} \eta}{\sqrt{2}} \right) \right]\ ,
    \end{aligned}
\end{equation}
where:
\begin{equation}
    \begin{aligned}
        g_{e, as} &= \sqrt{g^2 - (\Gamma+\mathrm{i}c_g\Delta)^2/8}\ ,\\
        \Gamma_{e, as} &= \Gamma + \mathrm{i}c_g{\Delta}\ .
    \end{aligned}
\end{equation}

The remained phonon spectrum density can be explained as:
\begin{equation}
    \begin{aligned}
    \kappa({\Delta},\eta) &= \frac{\langle \tilde{b}_{as}^\dagger({\Delta},\eta)\tilde{b}_{as}({\Delta},\eta)\rangle}{\langle \tilde{b}_{as}^\dagger({\Delta},0)\tilde{b}_{as}({\Delta},0)\rangle }\\
    &= \left|\mathbf{G}_{22}({\Delta}, \eta)\right|^2+ \Gamma \int_{0}^{t} \left|\mathbf{G}_{22}({\Delta}, \nu)\right| d \nu \ .
    \end{aligned}
\end{equation}
For the resonance case, when $\Delta=0$, we have:
\begin{equation}
    \kappa(\Delta=0,\eta) = 1-\mathrm{e}^{-\frac{\Gamma_{e,as}\eta}{2}}\frac{8g_{e,as}^2+\Gamma^2}{8g_{e,as}^2}\sin^2\left(\frac{g_{e,as}\eta}{\sqrt{2}}\right)\ .
\end{equation}

As an approximation when $g\gg \Gamma,\ g\gg |c{\Delta}|$, this result can be extended to the general case by taking the norm and replacing $\Gamma$ with $\Gamma_{e,as}$:
\begin{equation}
    \kappa({\Delta}, \eta)=1-\left|\mathrm{e}^{-\frac{\Gamma_{e,as}\eta}{2}}\frac{8g_{e,as}^2+\Gamma_{e,as}^2}{8g_{e,as}^2}\sin^2\left(\frac{g_{e,as}\eta}{\sqrt{2}}\right)\right| \ .
\end{equation}

\subsection{Detailed Derivation for Brillouin Entanglement}
\label{appendix:entangle}
In this section, we present the derivation process of the Brillouin interaction based optomechanics entanglement. The entangled pair generation can be achieved by down-conversion in quantum optics. In the optomechanis waveguide systems, the Hamiltonian for Stokes process also has a down-conversion like form, the only difference is the states it acts on which are one photon and one phonon, therefore the entangled pair it generated is a photon-phonon entangled pair.

The phenomenologically introduced effective Hamiltonian for anti-Stokes process reads (Eq.(\ref{Eq:Stokes_matrix})):
\begin{equation}
\begin{aligned}
\frac{\partial}{\partial \eta}
 \begin{pmatrix}
\tilde{a}_{s}(\Delta,\eta) \\ \tilde{b}_{s}^\dagger(\Delta,\eta)
\end{pmatrix}
=
\begin{pmatrix}
-\mathrm{i} c_g {\Delta}/2-\gamma/4 & - \mathrm{i} g/2\\
\mathrm{i} g & -\mathrm{i} c_g \Delta-\Gamma/2
\end{pmatrix}
\begin{pmatrix}
\tilde{a}_{s}(\Delta,\eta) \\ \tilde{b}_{s}^\dagger(\Delta,\eta)
\end{pmatrix}
+
\begin{pmatrix}
0 \\ \sqrt{\Gamma }\tilde{{\xi}}^\dagger
\end{pmatrix} \ .
\end{aligned}
\end{equation}
The thermal noise is introduced as the following:
\begin{equation}
    \begin{aligned}
&\left\langle \tilde{b}_{s}^\dagger(\Delta_1,\eta_1) \tilde{b}_{s}(\Delta_2,\eta_2)\right\rangle=n_{0} \delta(\Delta_1-\Delta_2)\delta(\eta_1-\eta_2) \ , \\
&\left\langle{\tilde{\xi}}^{\dagger}\left(\Delta_1,\eta_{1}\right) \tilde{{\xi}}\left(\Delta_2,\eta_{2}\right)\right\rangle=n_{th}\delta(\Delta_1-\Delta_2)\delta(\eta_1-\eta_2)\ . \\
\end{aligned}
\end{equation}.
The $n_0$ is the phonon number expectation at the initial state $\eta=0$, and the $n_{th}$ is the thermal phonon expecation number determined by the temperature of the enviornment. When the system is cooled by the cooling technique we proposed in the previous sections, $n_0<n_{th}$ can be achieved.

The quantum noise is introduced as the following:
\begin{equation}
    \begin{aligned}
&\left[ \tilde{b}_{s}(\Delta_1,\eta_1), \tilde{b}_{s}^\dagger(\Delta_2,\eta_2)\right] = \delta(\Delta_1-\Delta_2)\delta(\eta_1-\eta_2)\ , \\
&\left[\tilde{{\xi}}\left(\Delta_1,\eta_{1}\right), \tilde{{\xi}}^{\dagger}\left(\Delta_2,\eta_{2}\right)\right]=\delta(\Delta_1-\Delta_2)\delta(\eta_1-\eta_2)\ . \\
\end{aligned}
\end{equation}.

The $n_{0}$ refers to the average thermal phonon at phase-matching point at $t=0$. It is possible for $n_{0}<n_{th}$ when the system is pre-cooled by the laser cooling process, such as the coherent transfer-based Brillouin cooling we present in this paper. The exact expression for the matrix reads:
\begin{equation}
    \begin{aligned}
        \mathbf{G}_{11} &= \mathrm{e}^{-\frac{1}{4} \Gamma_{e, s}^* \eta-\mathrm{i}c_g\Delta \eta} \left[ \cosh \left( \frac{g_{e, s} \eta}{\sqrt{2}} \right) +\frac{\Gamma_{e, s}}{2\sqrt{2}g_{e, s}}\sinh \left( \frac{g_{e, s} \eta }{\sqrt{2}} \right) \right]\ ,\\
        \mathbf{G}_{12} &= -\mathrm{i}\mathrm{e}^{-\frac{1}{4}\Gamma_{e, s}^* \eta-\mathrm{i}c_g\Delta \eta}\frac{g}{\sqrt{2}g_{e, s}}  \sinh \left( \frac{g_{e, s} \eta}{\sqrt{2}} \right) \ ,\\
        \mathbf{G}_{21} &= \mathrm{i}\mathrm{e}^{-\frac{1}{4}\Gamma_{e, s}^* \eta-\mathrm{i}c_g\Delta \eta}\frac{\sqrt{2}g}{g_{e, s}}  \sinh \left( \frac{g_{e, s} \eta}{\sqrt{2}} \right) \ ,\\
        \mathbf{G}_{22} &= \mathrm{e}^{-\frac{1}{4} \Gamma_{e, s}^* \eta-\mathrm{i}c_g\Delta \eta} \left[ \cosh \left( \frac{g_{e, s} \eta}{\sqrt{2}} \right) -\frac{\Gamma_{e, s}}{2\sqrt{2}g_{e, s}}\sinh\left( \frac{g_{e, s} \eta }{\sqrt{2}}\right) \right] \ ,
    \end{aligned}
\end{equation}
where:
\begin{equation}
    \begin{aligned}
        g_{e,s} &= \sqrt{g^2 + (\Gamma+\mathrm{i} c_g \Delta)^2/8}\ ,\\
        \Gamma_{e,s} &= \Gamma + \mathrm{i} c_g \Delta\ .
    \end{aligned}
\end{equation}
The two mode EPR variables we choosed are:
\begin{equation}
    \begin{gathered}
u=\frac{1}{\alpha} X_{a}+ \alpha Y_{b}\ , \\
v=\frac{1}{\alpha} Y_{a}+\alpha X_{b}\ ,
\end{gathered}
\end{equation}
with $\alpha = 2^{-1/4}$. The $u,v$ can be written as:
\begin{equation}
    \begin{aligned}
    u&=2^{-1/2}\left[\left(2^{1/4}\tilde{a}_{s}+2^{-1/4}\mathrm{i}\tilde{b}_{s}^\dagger\right)+\textrm{h.c.}\right] \ ,\\
    v&=2^{-1/2}\left[-\mathrm{i}\left(2^{1/4}\tilde{a}_{s}+2^{-1/4}\mathrm{i}\tilde{b}_{s}^\dagger\right)+\textrm{h.c.}\right] \ .\\
\end{aligned}
\end{equation}
We have:
\begin{equation}
\begin{aligned}
    \quad\  2^{1/4}\tilde{a}_s(\Delta,\eta) + 2^{-1/4}\mathrm{i}\tilde{b}_{s}(\Delta,\eta)
    = & 2^{-1/4}\left(\sqrt{2}\mathbf{G}_{11}+\mathrm{i}\mathbf{G_{21}}\right)\tilde{a}_s(\Delta,0)
    + 2^{-1/4}\left(\sqrt{2}\mathbf{G}_{12}+\mathrm{i}\mathbf{G_{22}}\right)\tilde{b}_s^\dagger(\Delta,0)
    \\
    & \quad
    +2^{-1/4}\sqrt{\Gamma}\int_0^\eta\text{d}\nu\ \left(\sqrt{2}\mathbf{G}_{12}+\mathrm{i}\mathbf{G_{22}}\right) \tilde{\xi}^\dagger_b(\Delta,\nu)\ .\\
\end{aligned}
\end{equation}

The EPR variance is defined as :
\begin{equation}
        \sigma^2_{EPR}=\frac{\sigma^2 u+\sigma^2 v }{\alpha^2 + \alpha^{-2}}\ .
\end{equation}
For the entangled state, the Duan's criterion yields:
\begin{equation}
    \sigma^2_{EPR} < 1\ .
\end{equation}
Using the commutation relation, we obtained the $\Delta_{EPR}$ as:
\begin{equation}
     \sigma^2_{EPR}=\frac{1}{3}\left|\sqrt{2}\mathbf{G}_{11}+\mathrm{i}\mathbf{G}_{21}\right|^2 +\frac{1}{3}(2n_0+1) \left|\sqrt{2}\mathbf{G}_{12}+\mathrm{i}\mathbf{G}_{22}\right|^2+\frac{1}{3}\Gamma (2n_{th}+1)\int_0^\eta \text{d}\nu \left|\sqrt{2}\mathbf{G}_{12}+\mathrm{i}\mathbf{G}_{22}\right|^2
\end{equation}
We have:
\begin{equation}
    \begin{aligned}
    \sqrt{2}\mathbf{G}_{11} + \mathrm{i}\mathbf{G}_{21} &= \sqrt{2}\,  \mathrm{e}^{-\frac{1}{4} \Gamma_{e, s}^* \eta-\mathrm{i}c_g\Delta \eta}\left[\cosh\left(\frac{g_{e,s}\eta}{\sqrt{2}}\right) - \frac{g}{g_{e,s}}\sinh\left(\frac{g_{e,s} \eta}{\sqrt{2}}\right) + \frac{\Gamma_{e,s}}{2\sqrt{2}g_{e,s}}\sinh \left( \frac{ g_{e,s} \eta}{\sqrt{2}}\right) \right]\ ,\\
    \sqrt{2}\mathbf{G}_{12} + \mathrm{i}\mathbf{G}_{22} &= \mathrm{i}\mathrm{e}^{-\frac{1}{4}\Gamma_{e, s}^* \eta-\mathrm{i}c_g\Delta \eta}\left[\cosh\left(\frac{g_{e,s} \eta}{\sqrt{2}}\right)-\frac{g}{g_{e,s}}\sinh\left(\frac{g_{e,s} \eta }{\sqrt{2}}\right) - \frac{\Gamma_{e,s}}{2\sqrt{2}g_{e,s}}\sinh \left( \frac{ g_{e,s} \eta}{\sqrt{2}} \right) \right]\ .
    \end{aligned}
\end{equation}
When $g\gg \Gamma$, we have:
\begin{equation}
    \begin{aligned}
        \frac{\sqrt{2}}{2}\left|\sqrt{2}\mathbf{G}_{11} + \mathrm{i}\mathbf{G}_{21}\right| &\approx |\sqrt{2}\mathbf{G}_{12} + i\mathbf{G}_{22}|\\
        & \approx \mathrm{e}^{-\frac{1}{4} \Gamma_{e,s}^* \eta}\left| \mathrm{e}^{-\frac{g_{e,s} \eta}{\sqrt{2}}}-  \frac{\Gamma_{e,s}}{2\sqrt{2}g_{e,s}}\sinh \left( \frac{ g_{e,s} \eta }{\sqrt{2}}\right) \right|\ .
    \end{aligned}
\end{equation}
Therefore the EPR variance reads:
\begin{equation}
    \begin{aligned}
             \sigma^2_{EPR}&=\frac{1}{3}\left|\sqrt{2}\mathbf{G}_{11}+\mathrm{i}\mathbf{G}_{21}\right|^2 +\frac{1}{3}(2n_0+1) \left|\sqrt{2}\mathbf{G}_{12}+\mathrm{i}\mathbf{G}_{22}\right|^2+\frac{1}{3}\Gamma (2n_{th}+1)\int_0^\eta \text{d}\nu \left|\sqrt{2}\mathbf{G}_{12}+\mathrm{i}\mathbf{G}_{22}\right|^2 \\
             &\approx \left(\frac{2}{3}n_0+1\right)|G(t)|^2 +
             \frac{1}{3}(2n_{th}+1)\Gamma\int_0^\eta d\nu\  |G(\nu)|^2\ ,
    \end{aligned}
\end{equation}
where:
\begin{equation}
    G(\eta) = \mathrm{e}^{-\Gamma_{e,s}^* \eta/4}\left[ \mathrm{e}^{-\frac{g_{e,s} \eta}{\sqrt{2}}}-  \frac{\Gamma_{e,s}}{2\sqrt{2}g_{e,s}}\sinh \left( \frac{ g_{e,s} \eta }{\sqrt{2}}\right) \right]\ .
\end{equation}

\end{document}